\documentclass[manuscript,nonacm]{acmart}

\AtBeginDocument{%
  \providecommand\BibTeX{{%
    \normalfont B\kern-0.5em{\scshape i\kern-0.25em b}\kern-0.8em\TeX}}}

\usepackage{color, colortbl}
\usepackage{subcaption}

\usepackage{amsmath}
\newcommand{\NA}{---}

\begin{document}

\title{A Survey of Learned Indexes for the 
Multi-dimensional Space}

\author{Abdullah-Al-Mamun}
\affiliation{%
  \institution{Purdue University}
  \city{West Lafayette}
  \country{USA}
}
\email{mamuna@purdue.edu}

\author{Hao Wu}
\affiliation{%
  \institution{Keystone Strategy}
  \city{San Francisco}
  \country{USA}
  \postcode{Hao Wu contributed to this paper while studying at Purdue University}
 }
 \email{haowu3@alumni.cmu.edu}

\author{Qiyang He}
\affiliation{%
  \institution{Snowflake}
  \city{Bellevue}
 \country{USA}
 \postcode{Qiyang He contributed to this paper while studying at Purdue University}
  }
\email{qiyang.he@snowflake.com}

\author{Jianguo Wang}
\affiliation{%
  \institution{Purdue University}
  \city{West Lafayette}
  \country{USA}
  }
\email{csjgwang@purdue.edu}

\author{Walid G. Aref}
\affiliation{%
  \institution{Purdue University}
  \city{West Lafayette}
  \country{USA}
  }
\email{aref@purdue.edu}

\renewcommand{\shortauthors}{Abdullah-Al-Mamun, et al.}


\begin{abstract}
A recent research trend involves treating database index structures as Machine Learning (ML) models. In this domain, single or multiple ML models are trained to learn the mapping from keys to positions inside a data set. 
This class of indexes is known as ``Learned Indexes." Learned indexes have demonstrated improved search performance and reduced space requirements for one-dimensional data. The concept of one-dimensional learned indexes has naturally been extended to multi-dimensional (e.g., spatial) data, leading to the development of ``Learned Multi-dimensional Indexes". This survey focuses on learned multi-dimensional index structures. Specifically, it reviews the current state of this research area, explains the core concepts behind each proposed method, and classifies these methods based on several well-defined criteria. We present a taxonomy that classifies and categorizes each learned multi-dimensional index, and survey the existing literature on learned multi-dimensional indexes according to this taxonomy. Additionally, we present a timeline to illustrate the evolution of research on learned indexes. Finally, we highlight several open challenges and future research directions in this emerging and highly active field.
\end{abstract}

\maketitle

\section{Introduction}
Recently, due to tremendous progress in the field of machine learning, two research trends have emerged in the area of Database Systems (DB, for short): ML for DB, and DB for ML. In ML for DB, the goal is to develop ML-enhanced core database systems components, 
e.g.,
learned indexes~\cite{kraska2018case}, learned query optimizers~\cite{marcus2019neo}, and self-driving databases~\cite{pavlo2017self}. These initial learned systems components have demonstrated superior performance compared to their traditional counterparts. On the other hand, in DB for ML, the objective is to extend the traditional DB architecture, components, and query languages to support efficient in-database ML workloads~\cite{li2017mlog, jasny2020db4ml, chai2022data}.

\begin{figure}[tbp]
  \centering
  \includegraphics[width=0.8\textwidth]{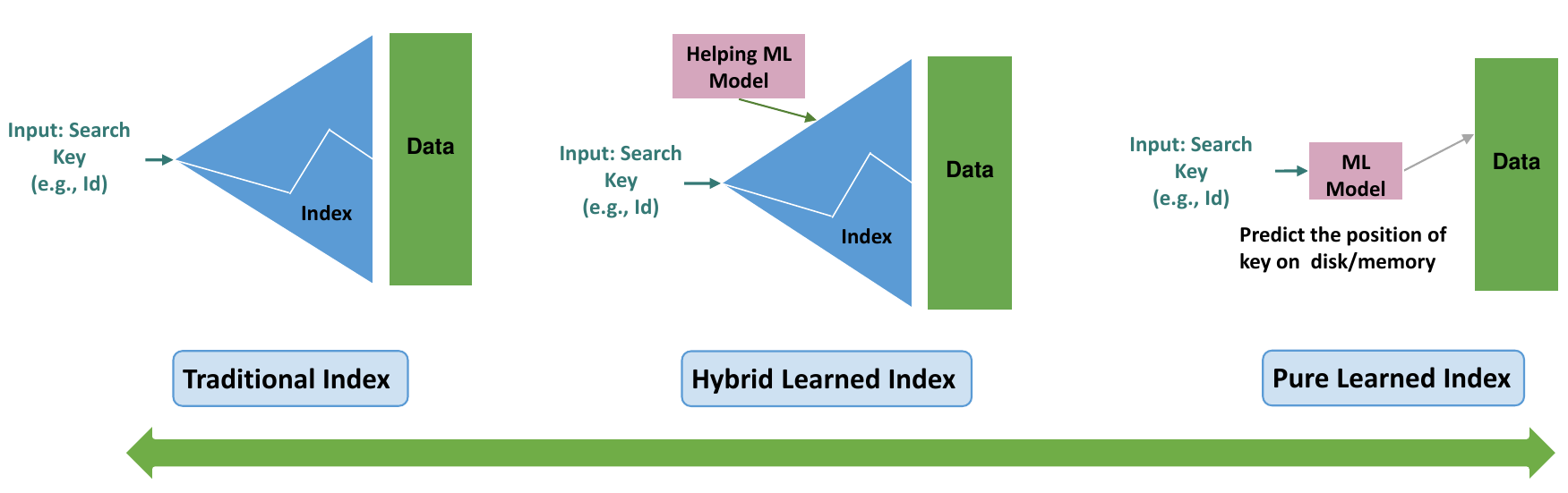}
  \caption{Spectrum of learned indexes}
  \label{fig:spectrum}

\end{figure}

This survey paper focuses on ML for DB, particularly on the idea of replacing traditional database index structures (e.g., B-tree~\cite{bayer1970organization,comer1979ubiquitous}) with ML models, first proposed in~\cite{kraska2018case}. The proposed Recursive Model Index (RMI, for short) can be considered as the first instance of a ``Learned Index". Note that traditional indexes provide theoretical guarantees on performance, are well-studied, and have been successfully integrated into practical data systems. In contrast, {\em pure} learned indexes learn the key-to-position mapping with some error-correction mechanisms to achieve better search performance, and reduce space requirements. On the other hand, there are {\em hybrid} learned indexes that optimize traditional indexes with helper ML models. The spectrum of learned indexes is given in Figure~\ref{fig:spectrum}.

Although the term ``Learned Indexes" has been coined very recently, the concept of using a learning mechanism in the context of database indexing has been studied previously. An example of an earlier index structure that combines ML techniques in the context of database indexing is the handwritten trie~\cite{arefHandWritten1995}. The handwritten trie employs Hidden Markov Models (HMM)~\cite{rabiner1986introduction} on a trie\cite{fredkin1960trie} structure to index the learned models. Notice that this handwritten trie focuses on the idea of {\em indexing the learned models} instead of {\em learning the index}. On the other hand, one of the earliest papers on a distribution-aware index structure for spatial (i.e., multi-dimensional) data can be found in~\cite{babu1997self}. In~\cite{babu1997self}, the proposed technique combines an R-tree~\cite{guttman1984r} with a self-organizing map~\cite{kohonen1990self}. Moreover, the initial promising results of one-dimensional learned indexes~\cite{kraska2018case} have inspired researchers to extend the concept in the context of multi-dimensional data. 
As a result, various methods for learned multi-dimensional indexes have been introduced in the recent years. 
One of the key assumptions in one-dimensional learned indexes is that the data can be sorted (i.e., totally ordered). However, there is no obvious total sort order for multi-dimensional data. As a result, it is challenging to define an error-correction mechanism in case of mis-predictions. Moreover, the layout of the multi-dimensional data might need to be re-arranged based on a pre-defined mechanism so that it can be easily learned by ML models. The choice of ML model might also vary from learned one-dimensional indexes due to the impact of dimensionality. In summary, learned multi-dimensional indexes need to 
address
additional research challenges 
that
are not associated with learned one-dimensional indexes. 
In this survey, we distinguish between multi- and high-dimensional data. In the context of multi-dimensional data, we assume that the dimension of the data is typically between $2$ to $10$. On the other hand, the dimension of the data can be very high (e.g., $100$) in the context of high-dimensional data. Although there are studies related to ML-enhanced high-dimensional indexes, we have not included them in this study unless some of the techniques in this domain have influenced the research of learned multi-dimensional indexes.

\begin{figure*}[tbp]
  \centering
  \includegraphics[width=\linewidth,]{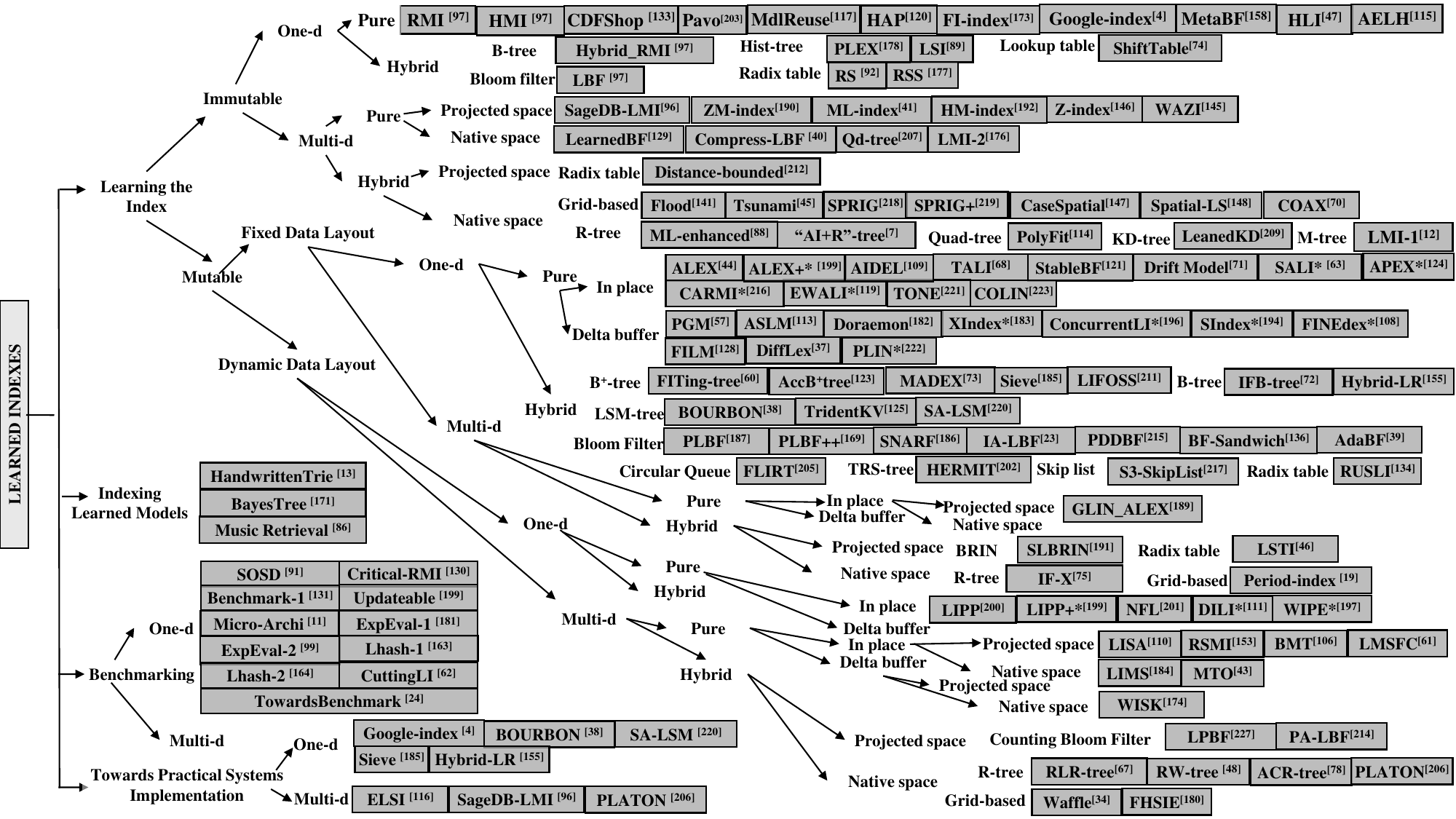}
  \caption{Taxonomy of learned indexes. The end of a branch indicates that there are no papers in that category as of the time this article has been written. An asterisk (*) symbol is used if the index natively supports concurrency. The hybrid learned indexes are categorized based on their underlying traditional data structures.}
\label{fig:taxonomy}
\end{figure*}

In this paper, we present a comprehensive survey of recent advances in the area of learned multi-dimensional indexes using that taxonomy given in Figure~\ref{fig:taxonomy}. In the taxonomy, we distinguish between two concepts: \textbf{indexing the learned models vs. learning the index}. By indexing the learned models, we refer to the following problem. Assume that we are given a collection of learned models, where each model can  recognize a certain object class, e.g., a learned model that recognizes the class car, another recognizes the class bike, etc. Given a query object, one needs to execute each of the models to identify the model that produces the highest recognition score. The problem can be formulated as {\em ``Can we index the learned models to accelerate the model identification process?''} In contrast, by {\em learning the index}, we refer to the problem of replacing a traditional database index structure with an ML model. For example, instead of searching a B$^+$-tree to locate the leaf page that contains an input search key, one uses an ML model that predicts the location that contains the search key.

\begin{figure*}[tbph] 
     \centering
     \begin{subfigure}[b]{0.60\linewidth}
         \centering
        \includegraphics[width=\linewidth]{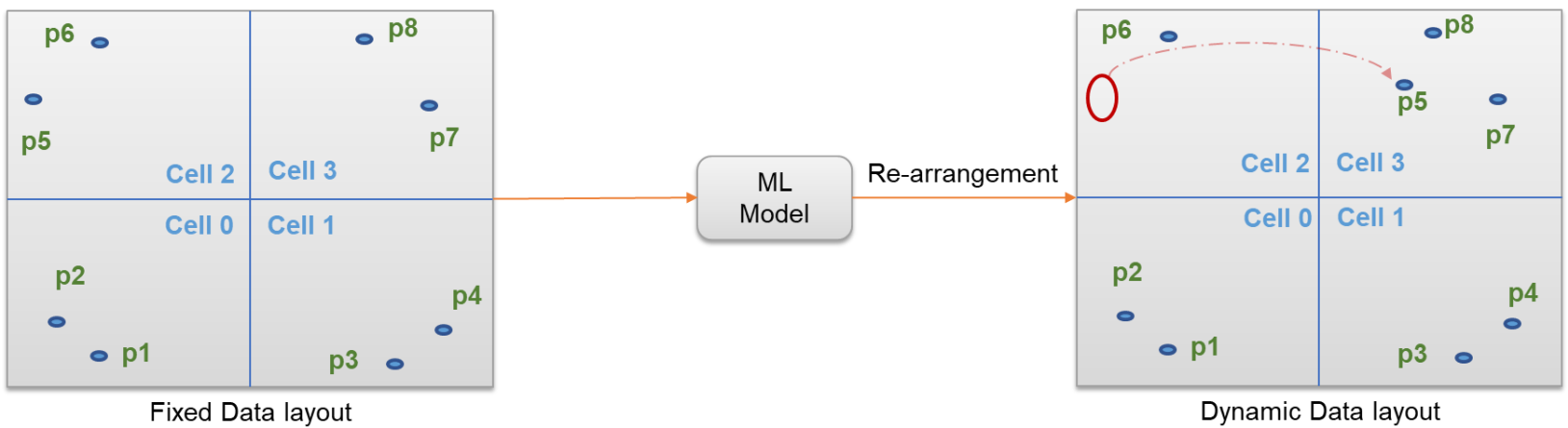}
        \caption{Fixed vs. Dynamic data layout}
        \label{subfig:Fixed vs. Dynamic data layout.}
     \end{subfigure}
     \hfill
     \begin{subfigure}[b]{0.60\linewidth}
         \centering
         \includegraphics[width=\linewidth]{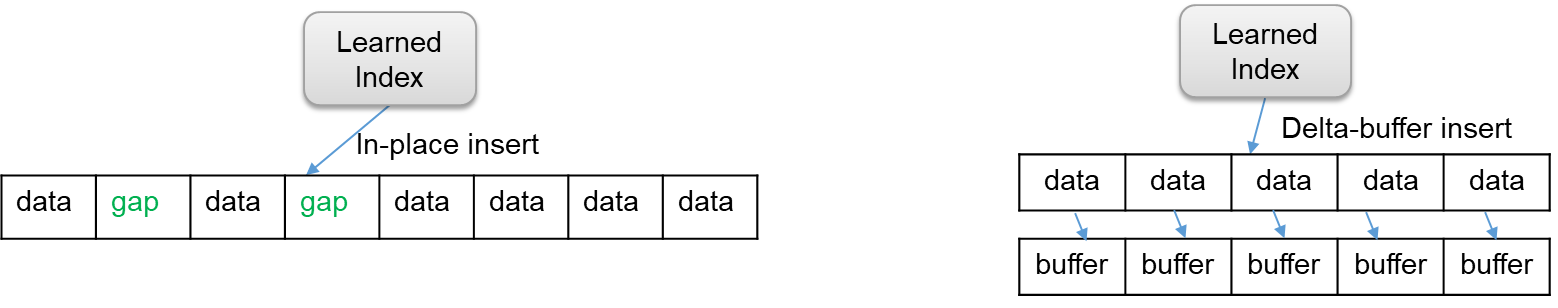}
        \caption{In-place vs. Delta buffer-based insertion}
        \label{subfig:InPlace_vs_deltabuffer}
     \end{subfigure}
     \hfill
     \begin{subfigure}[b]{0.50\linewidth}
         \centering
        \includegraphics[width=\linewidth]{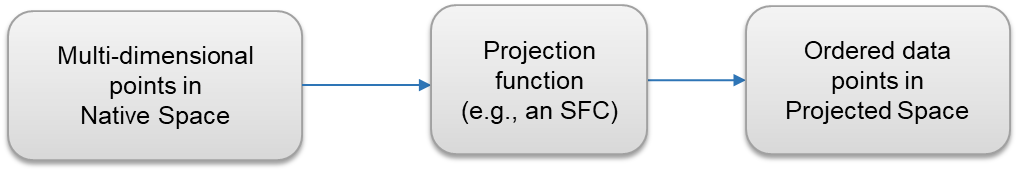}
        \caption{Native vs. Projected space}
         \label{subfig:Native_vs_projected}
     \end{subfigure}
        \caption{Illustration of the taxonomy criteria} 
        \label{fig:Taxonomy Criteria}
\end{figure*}

In the taxonomy, in the context of learning the index, we further distinguish among learned indexes along the following dimension: \textbf{learned indexes that support static datasets (i.e., Immutable) vs. learned indexes that support inserts/updates (i.e., Mutable)}. The issue of supporting static vs. dynamic datasets is crucial because learning an index requires offline training that is relatively slow in nature. Thus, learned indexes that support inserts/updates need to accommodate this fact, yet still offer online responses. 
In our taxonomy, mutable learned indexes are further divided into
\textbf{fixed vs. dynamic data layout}. A fixed data layout refers to the class of indexes where the layout of the data and the structure of the index are fixed before the index-building phase. On the other hand, if the layout of the data is arranged/re-arranged by the ML models while building the learned index, we refer to them as having a dynamic data layout. For example, in Figure~\ref{subfig:Fixed vs. Dynamic data layout.}, the initial fixed data layout is re-arranged using an ML model. 

Next, we distinguish between learned indexes along the dimensionality of the data: \textbf{learned indexes for one-dimensional data vs. learned indexes for multi-dimensional data}. 

Under each category, we further classify the indexes into \textbf{pure vs. hybrid learned indexes}. Pure learned indexes are designed without leveraging a traditional index structure (e.g., a B-tree or an R-tree). Moreover,  pure learned indexes are designed to replace a traditional index structure. In contrast, a hybrid learned index combines a traditional index structure with ML models to build an ML-enhanced index structure. As a result,  hybrid learned indexes are  categorized further based on their \textbf{underlying traditional data structure}.

On the other hand, mutable pure learned indexes are classified based on their policy for supporting new data insertions: \textbf{in-place vs. delta buffer insertion policies}. The in-place insertion policy uses ML models to find the position where the new data item should be inserted. Additionally, an in-place insertion policy might reserve some gaps in the index so that insert operations can be accommodated immediately. Conversely, the delta buffer insertion policy employs fixed-size buffers to hold the new data items for a short period of time. These temporary buffers are synchronized with the index structure periodically (e.g., during ML model re-training). However, we do not classify mutable hybrid indexes based on insertion strategies due to the presence of a traditional index, where that might influence the choice of the insertion strategy.
The difference between the two insertion policies are illustrated with the example given in Figure~\ref{subfig:InPlace_vs_deltabuffer}.

We further classify the learned multi-dimensional indexes into two classes: \textbf{indexes operating in the native space of the data vs. a projected space of the data}. Here, if the ML models are trained on the original representation of the multi-dimensional data, we refer to them as indexes built in native space. Conversely, if the multi-dimensional data is projected or linearized into a projected space using a projection function or a Space Filling Curve (SFC, for short)~\cite{sagan2012space, mokbel2003analysis}, we refer to them as indexes operating in a projected space (Figure~\ref{subfig:Native_vs_projected}).  

Additionally, in the taxonomy, we highlight studies related to benchmarking learned one-dimensional indexes.
At the time of writing this article,
we are not aware of any comprehensive benchmarking study in the area of learned multi-dimensional indexes. We also highlight and discuss studies related to the integration of learned indexes into practical systems.

\textbf{Contributions}. The main contributions of this survey are as follows. We provide an up-to-date coverage of learned multi-dimensional indexes (until the end of 2023). We present a taxonomy of existing learned indexes in both the one- and multi-dimensional spaces. We expect that 
upcoming learned index structures can be categorized within the taxonomy based on the proposed criteria.

\textbf{Paper Organization}. The rest of this paper is organized as follows: 
Section~\ref{section:Related Surveys} provides an overview of related surveys and tutorials. Section~\ref{section:Evolution} presents a timeline of the evolution of learned indexes. 
Section~\ref{section:Indexing the Learned Models} 
presents existing work in 
the area of indexing the learned models.
Sections~\ref{section:Learning_Immutable_one-d_multi-d},~\ref{section:Learning_Mutable_Fixed_one-d_multi-d}, and~\ref{section:Learning_Mutable_Dynamic_one-d_multi-d} present both one-dimensional and multi-dimensional indexes according to the taxonomy in Figure~\ref{fig:taxonomy}. 
The properties of immutable and mutable learned multi-dimensional indexes are summarized in Tables~\ref{table:multi-d_summary_Immutable},~\ref{table:fixed_layout_multi_d_summary_mutable}, and~\ref{table:multi-d_summary_mutable_dynamic}.
Section~\ref{section:List of ML Techniques} 
highlights the various 
ML techniques used in existing learned multi-dimensional indexes. 
Section~\ref{section:Benchmarking} discusses studies related to benchmarking 
learned indexes. Section~\ref{section:Towards Practical Systems} presents steps taken towards the integration of learned indexes in practical systems. 
Section~\ref{section:Open Challenges} provides an overview of open challenges and discusses future research directions. 
Finally, Section~\ref{section:Conclusion} concludes the paper.





\section{Related Surveys and Tutorials}~\label{section:Related Surveys}
The topic of learned indexes is relatively new,  and hence there are only a few related surveys in the literature. Most of the existing studies focus on learned one-dimensional indexes~\cite{ferragina2020learned, zhou2020database, liu2022survey}. Related conference-based tutorials on the subject of learned one-dimensional indexes can be found in~\cite{idreos2019auto,li2022machine, li2021ai, li2021machine, wang2016database}. Moreover, progress reports on ML in databases, and instance optimized data systems have been presented in~\cite{kraska2021ml, kraska2021towards}. A recent book on the topic of designing Data Structures with discussion on learned indexes for one-dimensional data can be found in~\cite{athanassoulis2023data}.

There is a 4-page conference tutorial on the subject of learned multi-dimensional indexes~\cite{al2020tutorial} that is an earlier version of this survey article by the same authors. 
That 
tutorial covers methods only until Year 2020. Moreover, the tutorial does not include  descriptions of the core concepts of the learned index structures listed in that short tutorial. This survey serves as an extension to that earlier conference tutorial and adds new criteria for classifying new indexes into a more sophisticated and comprehensive taxonomy as well as covers many more multi-dimensional learned indexes that have been developed until 2024.

Another related survey on multi-dimensional indexes can be found in~\cite{li2024survey}. However, in~\cite{li2024survey}, the survey focuses on the impact of modern hardware platforms (e.g., non-volatile memory, GPU), and ML algorithms in the context of multi-dimensional indexes.
There is also a series of related tutorials on big spatial data~\cite{sabek2019machine, sabek2020machine, sabek2021machine}. However, the focus of this series of tutorials is not on learned multi-dimensional index structures. There is also a related tutorial that covers ML-enhanced indexes particularly designed for high-dimensional data~\cite{echihabi2021new}. On the other hand, there are several comprehensive surveys for traditional multi-dimensional indexes~\cite{gaede1998multidimensional, samet1984quadtree, samet2006foundations}. As the topic of learned indexes is relatively new, to the best of our knowledge, this survey paper is the first to focus particularly on learned multi-dimensional index structures.

\section{Evolution of Learned Indexes}~\label{section:Evolution}

\begin{figure*}[h!]
  \centering
  \includegraphics[ width=\linewidth]{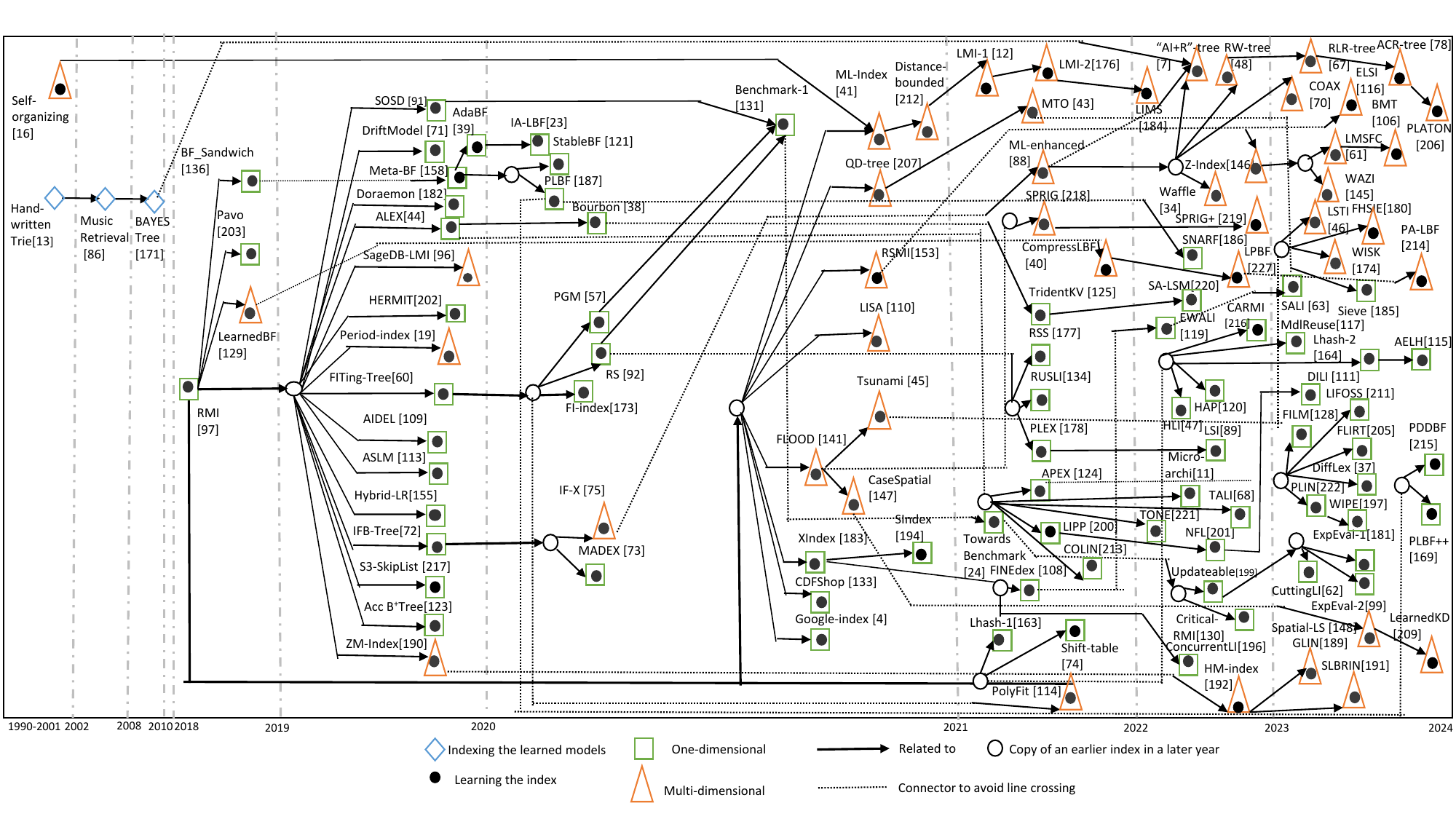}
  \caption{Timeline of the evolution of learned indexes. Lines connecting between the various learned multi-dimensional indexes reflect dependence of these indexes on earlier work.}
  \label{fig:timeline}
\end{figure*}

Figure~\ref{fig:timeline}, presents a chronological diagram to depict the evolution of both one- and multi-dimensional learned indexes. Here, we have grouped the papers on learned one- and multi-dimensional indexes based on their publication years. Moreover, we have used the $\rightarrow$ symbol to indicate if a later paper is related to an earlier one. The $\dashrightarrow$ is used when there is a line crossing. Additionally, a $\circ$ symbol is used to denote a copy of an earlier paper in a later year. One- and multi-dimensional indexes are differentiated using the $\color{green}\Box$ and $\color{orange}\triangle$ symbols, respectively.

\section{Indexing the Learned Models}~\label{section:Indexing the Learned Models}
Assume that we have a collection of learned models, where each model represents a certain class of objects, e.g., the class cat, dog, etc. Given an input instance, e.g., an image of an object, we need to execute each of the learned models to identify which model this input is likely to belong to, and hence identify the class of the input instance, e.g., being a cat with high probability. To avoid running each of the models on the instance, it would be good if we can {\em index} these learned models to distinguish between these models in a scalable way. We present this idea in the context of handwritten words and the 
Handwritten Trie~\cite{arefHandWritten1995} that highlight an example of the body of work for the class of Indexing the Learned Models. Other similar indexes in the same class can be found in~\cite{Aref2009}.


{\textbf{Handwritten Trie}~\cite{arefHandWritten1995}}
Given a collection of learned models, where each model corresponds to one handwritten word, e.g., a person's signature, handwritten indexes are used to speed up the search for a matching model given an input handwritten text, e.g., to help identify whose person this signature belongs to.
The core idea of the handwritten trie involves a trie structure as well as a collection of Hidden Markov Models (HMMs) that serve as the alphabet symbols in the trie nodes. The input to the handwritten trie is  handwritten text. This input is segmented into alphabet symbols, and the index is traversed by executing the HMM models associated with each of the trie node. Each HMM is trained to represent a pictogram class so that  each HMM accepts a specific pictogram with high probability. By traversing nodes at each level, the model chooses nodes with the highest probability for a particular letter (pictogram) and eventually obtains a set of nodes with the highest combined probability. 
Since the handwritten trie does not actually learn the distribution of key inputs, but rather indexes the HMM models, it is  categorized as an index for the  Learned Models rather than a learned index by itself.

Other indexes exist in this category. In the context of  music retrieval~\cite{jin2002indexing}, an R*-tree~\cite{beckmann1990r} is used to index the HMM models. In the BayesTree~\cite{seidl2009indexing}, an R-tree is employed to index a hierarchy of mixture density models. More indexes, e.g., as in~\cite{Aref2009}, fall under this category, but will not be discussed further in this survey.

\section{Learned Immutable Indexes}~\label{section:Learning_Immutable_one-d_multi-d}
Immutable learned indexes are built on static datasets and do not support dynamic inserts/updates.
In this section, we present the Learned Immutable Indexes in both the one- and multi-dimensional spaces. 

\subsection{The One-dimensional Case}
While one-dimensional learned indexes are not the focus of this survey, we present a few foundational indexes in the category as many indexes in the case of multi-dimensional learned indexes build on their one-dimensional counterparts.

\subsubsection{\textbf{Pure Learned Indexes}}
Here, we present the core concepts of RMI~\cite{kraska2018case} to highlight the class of Immutable One-dimensional Pure Learned Indexes.

\paragraph{{\textbf{RMI}~\cite{kraska2018case}}}
The key idea in~\cite{kraska2018case} is: ``Indexes are models." For example, given a key $k$, an index simply predicts the position of $k$ in a sorted array. Also, it has also been observed that ''a model predicting the position of a key within a sorted array effectively approximates the Cumulative Distribution Function (CDF)." As a result, indexes can be learned. 
In~\cite{kraska2018case}, it has been demonstrated how learned index structures can be implemented for three types of indexes: B-tree~\cite{bayer1970organization,comer1979ubiquitous}, Hash map~\cite{morris1968scatter, chi2017hashing}, and Bloom filter~\cite{bloom1970space}.
Similar to a B-tree, for range queries, the Recursive Model Index (RMI) structure is introduced that recursively chooses a model at 
multiple 
levels. With an input key, the root model chooses a child model based on its output. This continues until it reaches the leaf model
that 
predicts the actual position in the underlying sorted array. 

The prediction by the leaf-level ML model might not be accurate. As a result, an error correction mechanism is employed to correct the misprediction within a predefined error bound (i.e., min and max error). 
Because data in the one-dimensional case is sorted, and hence a total order exists among the elements of the indexed array, any misprediction can be corrected by searching either to the left or to the right of the mispredicted value.   A binary or exponential search operation can be performed to search for the correct value from the mispredicted location. The proposed RMI structure can be built using a mixture of ML models (e.g., neural networks~\cite{aggarwal2018neural} or linear regression~\cite{james2023linear}). Moreover, at the bottom level of RMI, a traditional B-tree can also be used as a model, introducing the idea of Hybrid Learned Indexes (i.e., Hybrid RMI).
For point queries, the idea of Hash-Model Index (HMI) has been proposed in the same paper. For minimizing collisions, HMI leverages the CDF of the dataset by employing learned hash functions.
For existence indexes, the Learned Bloom Filter (LBF) has been introduced~\cite{kraska2018case}. LBF is formulated as a classification problem. It proposes adopting a hybrid approach by integrating a traditional bloom filter with an ML model. During query processing, the key is first passed into a learned model to decide on existence, and a traditional bloom filter is used to catch any false negative cases from the learned model.

Other example indexes in this category are as follows. CDFShop~\cite{CDFShop} is a tool that guides users on tuning the parameters of the RMI index structure. Pavo~\cite{xiang2018pavo} leverages Recurrent Neural Networks (RNN)~\cite{cho2014learning} to replace the hash function in a traditional inverted index structure. In MdlReuse~\cite{liu2023efficient}, pre-trained ML models are used instead of training a model during the construction of the learned index. HAP~\cite{liu2022hap} is a Hamming space indexing framework that leverages ML models for cost estimation, query processing, and index compression. The FI-index~\cite{setiawan2020function} investigates the use of function interpolation models as an alternative choice of ML model in the context of learned indexes. In Google-index~\cite{abu2020learned}, a learned index is introduced for a disk-based distributed system. In MetaBF~\cite{rae2019meta}, a learned bloom filter is proposed by leveraging meta-learning~\cite{santoro2016meta}. HLI~\cite{ding2022error} uses ML models with error bounds for leaf nodes and models without error bounds for inner nodes to achieve the benefits of both approaches. AELH~\cite{lin2023learning} is a learned hash index structure that leverages Autoencoders~\cite{vincent2008extracting} for binary hash codes.


\subsubsection{\textbf{Hybrid Learned Indexes}}
Hybrid learned indexes combine traditional index structures with ML models to build ML-enhanced
index structures. In this section, we present the core concepts of RS~\cite{kipf2020radixspline} as an example to highlight the class of Immutable One-dimensional Hybrid Learned Indexes. Then, in the section to follow, we survey in more detail, the multi-dimensional case.

\paragraph{{\textbf{RS}~\cite{kipf2020radixspline}}}
RadixSpline (RS, for short) is a hybrid learned index structure that combines the linear spline model with a traditional radix table~\cite{leis2013adaptive}. Given a search key, the traditional radix table is used to locate the two spline points bounding the search key. RS can be constructed in a single pass if the data is sorted. The steps to construct the RS index are as follows: i)~Fit a linear spline to the CDF of the data to ensure a certain error-bound, and collect a set of spline points, and ii)~Construct an approximate index of the spline points using a radix table. The proposed index supports both equality and range predicates. However, performance of the RS index  can be negatively impacted by skewed data distributions. To address this, a tree-structured radix table can be utilized. Note that the RS structure can be tuned using only two hyperparameters for a given data and memory budget.

The other indexes in this category are as follows. Hybrid\_RMI~\cite{kraska2018case} combines a traditional B-tree with an RMI structure to create a hybrid index structure. PLEX~\cite{stoian2021plex} leverages a traditional Hist-tree~\cite{crotty2021hist} to build a learned index with a single tunable parameter. LSI~\cite{kipf2022lsi} also leverages a traditional compact Hist-tree to construct a learned index for unsorted data. ShiftTable~\cite{hadian2021shift} is a helper error correction layer that can be combined with the ML models of a learned index. LBF~\cite{kraska2018case} introduces the learned bloom filter structure by combining ML models with traditional bloom filters. RSS~\cite{spector2021bounding} extends the methods of RS~\cite{kipf2020radixspline} for indexing strings.

\subsection{The Multi-dimensional Case}~\label{subsection:Learning_Immutable_multi-d}

In this section, we explain the learned multi-dimensional indexes along the various dimensions of our taxonomy. The properties of the class of immutable learned multi-dimensional indexes are presented in Table~\ref{table:multi-d_summary_Immutable}. We highlight the types of queries supported by each of the learned multi-dimensional indexes. For point queries, some indexes do not explicitly provide a query processing algorithm or experimental results. However, if an index can be easily extended to support point queries, we highlight that by listing that the index supports point query processing.  
On the other hand, due to nature of the application domain, an index might be designed emphasizing efficiency over accuracy. As a result, an index can support either exact or approximate query processing. Here, approximate query processing refers to the event of missing some results from the set of exact answers to the query. We indicate whether a supported query type returns an exact or an approximate answer.
Notice that we have excluded the learned multi-dimensional bloom filters (e.g., LearnedBF~\cite{macke2018lifting}, CompressLBF~\cite{davitkova2021compressing}) in the context of supported query types because they are considered as probabilistic existence index structures. 

\begin{table}[h!]
\caption{Properties of the Immutable Learned Multi-dimensional Indexes} 
\centering 
\begin{tabular}{p{0.2\linewidth}p{0.05\linewidth}p{0.1\linewidth}p{0.09\linewidth}p{0.07\linewidth}p{0.07\linewidth}p{0.07\linewidth}p{0.07\linewidth}} 
\toprule
Index&Pure Learned&Hybrid Learned&Data Space&Point Query&Range Query& kNN Query &Join Query\\
\midrule
SageDB-LMI~\cite{kraska2019sagedb} & $\checkmark$ & $\times$ & Projected & Exact & Exact & $\times$ & $\times$\\
\hline
ZM-index~\cite{wang2019learned} & $\checkmark$ & $\times$ & Projected & Exact & Exact & $\times$ & $\times$\\
\hline
ML-index~\cite{davitkova2020ml} & $\checkmark$ & $\times$ & Projected & Exact & Exact & Exact & $\times$\\
\hline
HM-index~\cite{wang2021spatial} & $\checkmark$ & $\times$ & Projected& Exact & Exact & $\times$ & $\times$\\
\hline
Z-index~\cite{pai2022towards} & $\checkmark$ & $\times$ & Projected & Exact & Exact & $\times$ & $\times$\\
\hline
WAZI~\cite{pai2023workload} & $\checkmark$ & $\times$ & Projected & Exact & Exact & $\times$ & $\times$\\
\hline
LearnedBF~\cite{macke2018lifting} & $\checkmark$ & $\times$ & Native & \NA & \NA & \NA & \NA\\
\hline
CompressLBF~\cite{davitkova2021compressing} & $\checkmark$ & $\times$ & Native & \NA & \NA & \NA & \NA\\
\hline
Qd-tree~\cite{yang2020qdtree} & $\checkmark$ & $\times$ & Native & Exact & Exact & $\times$ & $\times$\\
\hline
LMI-2~\cite{slaninakova2021data} & $\checkmark$ & $\times$ & Native & Approx.& Approx. &Approx. & $\times$\\
\hline
Distance-bounded~\cite{ZacharatouKSPDM21} & $\times$ & Radix table & Projected & Approx.& Approx. & $\times$ & Approx.\\
\hline
Flood~\cite{nathan2020learning} & $\times$ & Grid & Native & Exact & Exact & $\times$ & $\times$\\
\hline
Tsunami~\cite{DingNAK2020tsunami} & $\times$ & Grid & Native& Exact & Exact & $\times$ & $\times$\\
\hline
SPRIG~\cite{Zhang2021sprig} & $\times$ & Grid & Native & Exact & Exact & Exact& $\times$\\
\hline
SPRIG+~\cite{zhang2022efficient} & $\times$ & Grid & Native & Exact & Exact & Exact & $\times$\\
\hline
ML-enhanced~\cite{kang2021mlenhanced} & $\times$ & R-tree & Native & $\times$ & $\times$ & Approx.& $\times$\\
\hline
``AI + R''-tree~\cite{mamun2022ai+} & $\times$ & R-tree & Native & Exact & Exact & $\times$ & $\times$\\
\hline
CaseSpatial~\cite{pandey2020case} & $\times$ & Grid & Native & Exact & Exact & $\times$ & $\times$\\
\hline
Spatial-LS~\cite{pandey2023enhancing} & $\times$ & Grid & Native & Exact & Exact & $\times$ & Exact\\
\hline
COAX~\cite{hadian2023coax} & $\times$ & Grid & Native & Exact & Exact & $\times$ & $\times$\\
\hline
PolyFit~\cite{li2021polyfit} & $\times$ & Quad-tree & Native &  Approx. & Approx. & $\times$ & $\times$\\
\hline
LearnedKD~\cite{yongxin2020study} & $\times$ & KD-tree & Native & $\times$ & $\times$ & Approx.& $\times$\\
\hline
LMI-1~\cite{antol2021learned} & $\times$ & M-tree & Native & Approx.& Approx. &Approx. & $\times$\\

\bottomrule
\end{tabular}
\label{table:multi-d_summary_Immutable}
\end{table}

\subsubsection{\textbf{Pure Learned Indexes in the Projected Space}}
\paragraph{{\textbf{SageDB-LMI}~\cite{kraska2019sagedb}}}
SageDB is an ML-enhanced database system that has been envisioned in~\cite{kraska2019sagedb}. SageDB includes a learned multi-dimensional index (SageDB-LMI, for short) that is one of the earliest attempts to extend techniques similar to RMI~\cite{kraska2018case} in the context of multi-dimensional data. 
SageDB-LMI projects the multi-dimensional data points into one-dimensional space by consecutively sorting and partitioning the points along a sequence of dimensions (e.g., first x-dimension then y-dimension) into uniformly-sized cells.
This produces a layout that is easily learnable compared to the more complex space-filling curves, e.g., the  Z-order SFC~\cite{mokbel2003analysis}. Notably, any one-dimensional learned indexing method can be applied to the projected space. Consequently, in the second step, 
SageDB-LMI
uses a trained CDF model (e.g., RMI) to predict the physical location of the point. In an in-memory, read-only experimental setup, SageDB-LMI   outperforms a traditional R-tree in terms of average query time and index size.

\paragraph{{\textbf{ZM-index}~\cite{wang2019learned}}}
Similar to SageDB-LMI~\cite{kraska2019sagedb}, the ZM-index projects multi-dimensional spatial data into a one-dimensional projected space. Particularly, the ZM-index involves two essential components: 
1. A Z-order~\cite{orenstein1984class, morton1966computer, Peano1890} curve to linearize the multi-dimensional space, and 2. A multi-staged model index to support search. The Z-order curve sorts the multi-dimensional data according to the corresponding Z-addresses computed through bit interleaving. The multi-staged model index consists of neural network models at multiple stages
that
recursively partition the data space into sub-regions. Similar to RMI~\cite{kraska2018case}, given a query key, the model at Stage $i$ predicts a position based on the CDF and the number of all keys. After that, it chooses a model at Stage $i+1$ according to the predicted position or directly outputs the predicted position if already at the leaf level. The steps for range query processing using ZM-index are as follows: (i)~Computing Z-addresses of the start and end points of the range query. (ii)~Predicting the positions of the start and end points using the multi-staged model index structure. (iii)~Finding the exact positions of the start and end points using model-based search. (iv)~Scanning through the points within the range. Although the ZM-index can achieve query processing time similar to an R-tree, it can significantly reduce the index size compared to its traditional counterpart.

\paragraph{{\textbf{ML-index}~\cite{davitkova2020ml}}}
In the case of learned multi-dimensional indexes in the projected space, it is desirable for the multi-dimensional data to be projected in an order that can be easily learned by ML models. To achieve this goal, the proposed Multi-dimensional Learned
index
(ML-index, for short) generalizes the idea of the previously proposed iDistance~\cite{jagadish2005idistance}. The ML-index partitions and transforms the data into one-dimensional values based on distribution-aware reference points. It proposes an efficient scaling method so that, after projecting the multi-dimensional points into the one-dimensional space, the spatial proximity in the native space is well-preserved in the lower dimension. Similar to the learned B-tree index
in~\cite{kraska2018case}, a recursive model index is applied to the projected space. The ML-index can support point, range, and kNN queries.

\paragraph{{\textbf{HM-index}~\cite{wang2021spatial}}}
The Hilbert Model index (HM-index, for short) follows the similar principle that has been used for the ZM-index~\cite{wang2019learned}. Here, the core idea is to project the multi-dimensional data into the one-dimensional space using the Hilbert SFC~\cite{Hilbert1891, sagan2012space} 
and applying one-dimensional learned indexing techniques on the projected space. Particularly, a two-stage model has been used for learning the underlying data distribution. During the prediction step, given a query point, the model either predicts the position of the query point or a range in which the position of the query points can be searched. Moreover, due to the application of the Hilbert linear ordering, an error correction mechanism can be used 
in case
of an incorrect prediction with an error bound. Notice that the nearby data points in the projected Hilbert space may be far apart in the native space. As a result, it can impact the performance of range query processing. 
Thus, a query partitioning technique 
based on n-order Hilbert regions 
has been adopted to minimize the impact on performance.

\paragraph{{\textbf{Z-index}~\cite{pai2022towards}}}
Many spatial indexes use a pre-defined SFC, 
e.g., the Z-curve or Hilbert-curve, to order multi-dimensional data. However, existing SFCs are not designed to be instance-optimized~\cite{kraska2021towards} for given data and query workloads. The idea of an instance-optimized SFC-based index has been investigated~\cite{pai2022towards}. Particularly, for a given data and query workloads, an instance-optimized variant of the Z-index has been proposed by varying the partitioning and the ordering~\cite{ramakrishnan2003database}. The goal is to reduce the number of false positives during range query processing. In the proposed algorithm, two heuristic-based approaches are explored: the Independence-based Heuristic, and the Sampling-based Heuristic. For the sampling-based heuristic, after the partitioning step, a learned density model is applied to approximately calculate the number of data points falling into each of the children-cells. Notice that an instance-optimized SFC is directly applicable to the class of learned multi-dimensional indexes operating on a projected space.

\paragraph{{\textbf{WAZI}~\cite{pai2023workload}}}
The Workload-aware Z-Index (WAZI, for short) is an extension of the previously proposed instance-optimized Z-index~\cite{pai2022towards}. Given a spatial dataset and range query workloads, the proposed index optimizes the partitioning and Z-ordering. WAZI's partitioning technique produces cells that are accessed by a similar set of range queries. This enables the proposed index to reduce extraneous cell accesses during query processing. The benefit will persist with the proposed partitioning and ordering as long as the distribution of the query workload remains unchanged. During the index construction phase, a greedy algorithm is proposed for the partitioning and ordering of the cells. After the index construction phase, a particular partitioning and ordering are chosen by minimizing a pre-defined objective function based on the number of accessed data points. Here, Random Forest Density Estimation (RFDE)~\cite{wen2022random} models are used to approximate the exact data and query distributions.

\subsubsection{\textbf{Pure Learned Indexes in Native Space}}
\paragraph{{\textbf{LearnedBF}~\cite{macke2018lifting}}}
LearnedBF extends the concept of Learned Bloom Filter (LBF)~\cite{kraska2018case} to design a learned multi-dimensional bloom filter. For designing LearnedBF, firstly, the strings with k-tuples are converted into an embedding vector. Notice that Recurrent Neural Networks (RNN) with gated recurrent units~\cite{cho2014learning}, and direct embedding technique are used for high- and low-cardinality attributes, respectively. 
Moreover, it has been observed that a learned bloom filter can handle multidimensional data (e.g., k-tuple) effectively if there is a ``co-occurrence structure" between in-index k-tuples that are different from out-of-index tuples. Based on the observation, the embedding vectors are concatenated and sent as an input to a densely connected neural network layer. The output of this layer is used as an input to a sigmoid function. After that, a sandwich structure~\cite{BF-sandwich10.5555/3326943.3326986} is employed to get the final output of the LearnedBF. 

\paragraph{{\textbf{CompressLBF}~\cite{davitkova2021compressing}}}
CompressLBF has been 
introduced 
to reduce the space consumption of the previously proposed LearnedBF~\cite{macke2018lifting}. It is observed that the size of the trained ML model is significantly impacted by the number of distinct values in the input. Particularly, for a column with a high number of unique values, the size of the corresponding embedding matrix will grow linearly. As a result, it has been proposed to compress the input embedding by splitting a column into several sub-columns where the number of sub-columns is pre-defined. This split operation creates sub-columns with fewer dimensions. As a result, the embedding of the input will be smaller in size, and hence the proposed filter requires less space. Moreover, the compressed filter achieves high accuracy and saves model training time.

\paragraph{{\textbf{Qd-tree}~\cite{yang2020qdtree}}}
It has been observed that modern big data analytical systems partition data mainly by two approaches: i)~Hash/time-based, and ii)~Clustering-based~~\cite{jain1999data}. These existing methods do not consider the distribution of the query workload while partitioning the data. Given a particular data and query workloads, the Query-data Routing Tree (Qd-tree, for short) leverages Reinforcement Learning~\cite{kaelbling1996reinforcement} (RL, for short) to partition the data so that the number of blocks accessed by a particular query workload is minimized. The Qd-tree can be considered as a workload-aware multi-dimensional index structure where each non-leaf node partitions the data using a particular query predicate. Moreover, the data in a leaf node is routed to the same disk block. For the formulation of the Qd-tree construction process as a Markov Decision Process (MDP)~\cite{puterman1990markov}, the set of nodes are represented as states, the action space is represented as the set of query predicates (i.e., allowed cuts), and the reward is calculated using the number of skipped blocks over all queries. Notice that calculating the actual reward (i.e., number of skipped blocks) by executing queries is a costly process. As a result, a sampling method has been leveraged to avoid the costly query execution. On the other hand, the proposed RL agent, namely ``Woodblock", consists mainly of two learnable networks: i)~Policy network, and ii)~Value network. Moreover, Proximal Policy Optimization~\cite{schulman2017proximal} (PPO, for short) is used as the underlying learning algorithm.

\paragraph{{\textbf{LMI-2}~\cite{slaninakova2021data}}}
LMI-1~\cite{antol2021learned} is a Learned Metric Index that leverages predictions from a hierarchical structure with ML models for query processing. Notice that LMI-1 is a hybrid structure that requires a pre-existing index structure to partition the data. LMI-2 extends the LMI-1 by eliminating the requirement of a pre-existing index structure to partition the data objects. Moreover, LMI-2 adopts an unsupervised approach for ML model training by leveraging clustering~\cite{jain1999data} techniques. As a result, LMI-2 requires less index construction time than LMI-1 and outperforms LMI-1 in terms of query processing time.

\subsubsection{\textbf{Hybrid Learned Indexes in the Projected Space}}
\paragraph{{\textbf{Distance-bounded}~\cite{ZacharatouKSPDM21}}}
Most traditional spatial data processing methods follow a pipeline of coarse-grained filtering followed by exact geometric tests. In contrast, the method proposed in~\cite{ZacharatouKSPDM21} 
advocates for fine-grained raster approximations to eliminate the need for exact geometric tests, prioritizing efficiency over accuracy. As a result, the proposed method leverages a raster-based approximation of spatial objects for approximate query processing. Moreover, a user-defined  distance bound (i.e., error bound) is used to control the accuracy of the spatial approximation. Based on the user defined distance bound, the space is divided into uniform two-dimensional raster cells. Subsequently, the two-dimensional raster cells are projected into a one-dimensional array using a SFC. After that a one-dimensional learned index (e.g., RadixSpline~\cite{kipf2020radixspline}) is constructed to learn the position of the cells in the one-dimensional array. Notice that the learned index is used for approximate point query processing. On the other hand, to enhance query performance for polygons, an Adaptive Cell Trie structure (ACT, for short)~\cite{kipf2020adaptive} is used.


\subsubsection{\textbf{Hybrid Learned Indexes in Native Space}}
\paragraph{{\textbf{Flood}~\cite{nathan2020learning}}}
Flood is a clustered in-memory read-only learned multi-dimensional index (for column stores) optimized for specific datasets and query workloads. During the index construction phase, Flood takes query workloads as input and learns to automatically create a data layout optimized for the given query distribution. Given a dataset with d dimensions, Flood models the empirical Cumulative Distribution Function (CDF) of each dimension using RMI~\cite{kraska2018case}. Then, Flood 
uses these models to create partitions of equal size, ensuring each partition contains an equal number of data points. These partitions are grouped to form a grid structure. Flood optimizes the data layout for particular workloads by adjusting the number of partitions in each dimension. It uses d-1 dimensions for partitioning, designating the last dimension as the ``sort dimension". Flood fine-tunes the parameters of its grid structure using a cost model, 
and applies 
a gradient-descent algorithm to the cost model to optimize its parameters jointly. During query processing, for each query, Flood identifies all grid cells intersected by the query, and follows a projection-refinement-scan approach.

\paragraph{{\textbf{Tsunami}~\cite{DingNAK2020tsunami}}}
Traditional multi-dimensional indexes, 
e.g., 
the $k$-d tree~\cite{Bentley1975kdtree}, partition the underlying space based on the data distribution alone. In contrast, learned indexes,
e.g., 
Flood~\cite{nathan2020learning}, optimize for both data and query workloads. However, Flood's performance suffers under skewed query workloads and correlated data. To overcome Flood's limitations, Tsunami is proposed. Tsunami introduces a Grid Tree structure and an Augmented Grid to address the issues of skewed query workloads and correlated data, respectively. The proposed grid tree is a lightweight Decision Tree~\cite{quinlan1986induction}, and the augmented grid uses conditional CDFs and functional mappings. For the grid tree, Tsunami clusters queries by selectivity and builds a Grid Tree for each query type. Each node in the Grid Tree is split into several ranges along one dimension (except leaf nodes) until it reaches a minimum threshold or exhibits low query skew. For the augmented grid structure, Tsunami employs three strategies: i)~Partitioning Dimension X independently by its CDF (similar to Flood), ii)~Eliminating Dimension X by constructing a mapping from X to Y (if they are monotonically correlated), and iii)~Partitioning X dependent on Dimension Y by CDF(X|Y). To find the best augmented grid structure, Tsunami defines the number of partitions and a search strategy including the above three approaches. Then, it 
uses adaptive gradient descent to iteratively search for an optimal strategy. Tsunami outperforms the previously proposed Flood in scenarios with skewed query workloads and correlated data.

\paragraph{{\textbf{SPRIG}~\cite{Zhang2021sprig}}}
Spatial Interpolation Function based Grid (SPRIG, for short) is a grid-based learned multi-dimensional index designed for read-only workloads. SPRIG applies a sampling technique to the dataset, and uses the sampled data to build an adaptive grid structure. Then, it 
leverages the sampled data to fit a spatial interpolation function~\cite{mitas1999spatial}, specifically using the bilinear interpolation function. During query processing, given a search key, SPRIG predicts the approximate position of the key using the learned interpolation function. Notice that SPRIG requires a local search as an error correction mechanism following the prediction step. Moreover, to provide an error bound, SPRIG introduces a maximum estimation error calculated using the query workload. SPRIG can process both range and kNN queries.

\paragraph{{\textbf{SPRIG+}~\cite{zhang2022efficient}}}
Observing the imprecision of a single spatial interpolation function and the large prediction error in SPRIG~\cite{Zhang2021sprig}, SPRIG+ combines the idea of space-partitioning trees with the prediction mechanism. Firstly, SPRIG+ divides the two-dimensional grids into four sub-regions recursively. Next, one spatial interpolation function is learned for each region. Furthermore, to optimize storage space, SPRIG+ stores only the integer coordinates (used for calculating the interpolation function coefficients) instead of keeping the double coefficients (for lazy calculation). Additionally, SPRIG+ compresses all data into a bit vector, ensuring that the number of bits occupied by each integer remains as small as possible.

\paragraph{{\textbf{CaseSpatial}~\cite{pandey2020case}}}
The techniques proposed in Flood~\cite{nathan2020learning} have been applied to five in-memory traditional spatial index structures in CaseSpatial~\cite{pandey2020case}: Fixed-grid~\cite{bentley1979data}, Adaptive-grid~\cite{nievergelt1984grid}, k-d Tree~\cite{Bentley1975kdtree}, Quad-tree~\cite{samet1984quadtree}, and STR-tree~\cite{leutenegger1997str}. Moreover, the query processing in all 
these
techniques is 
performed 
in three steps: Index lookup, Refinement, and Scan. Similar to Flood, data in each partition are sorted using one dimension. During range query processing, while searching within a partition, it is proposed to use a one-dimensional learned index (e.g., RadiXSpline~\cite{kipf2020radixspline}) on the sorted dimension instead of a binary search. As a result, performance has improved for range queries with low selectivity. However, there is less improvement in the case of range queries with high selectivity. This has led to the conclusion that the performance of a range query with low selectivity can be significantly improved by employing learned indexing techniques. It has also been observed that filtering in one dimension and refining in the other dimension using a one-dimensional learned index can outperform methods that filter on two dimensions.

\paragraph{{\textbf{Spatial-LS}~\cite{pandey2023enhancing}}}
Spatial-LS is an extension of the techniques proposed in  CaseSpatial~\cite{pandey2020case}. Here, six different traditional spatial partitioning techniques have been considered for experiments, and these techniques are applied to achieve instance-optimization. Notice that in CaseSpatial~\cite{pandey2020case}, only the range query is supported. On the other hand, Spatial-LS 
supports 
point, range, distance, and spatial join queries. Moreover, 
their extensive experimental studies 
show that: i)~Properly tuned grid-based structures can outperform tree-based structures due to fewer random accesses and the benefit of learned search within larger partitions; ii)~Within a large partition, a learned model performs better than binary search; and iii)~The impact of ML-enhanced grid-based index structures is less in the case of queries with high selectivity. Notice that all these techniques are developed considering an in-memory setup. As a result, the benefits of the proposed techniques might not be directly applicable in a disk-based setup due to the diminished importance of searching within each partition.

\paragraph{{\textbf{COAX}~\cite{hadian2023coax}}}
It has been observed that two or more attributes are correlated in many real-world multi-dimensional datasets. As a result, these correlations can be leveraged for dimensionality reduction. Correlation-Aware Indexing (COAX, for short) learns the correlations 
among the 
attributes. The main idea is to construct a multi-dimensional index by excluding the highly-correlated attributes. As a result, the constructed index will be significantly reduced in size, 
and that 
can enhance query processing performance. Moreover, if a query involves a non-indexed attribute,
say $a1$, 
that is correlated with another indexed attribute $a2$, COAX uses only the correlated indexed attribute to process the query. COAX learns the correlation by learning Soft Functional Dependency (softFD)~\cite{ilyas2004cords}. Furthermore, COAX is implemented with Grid Files~\cite{nievergelt1984grid} by adopting a hybrid approach and supports point and range queries.

\paragraph{{\textbf{ML-enhanced}~\cite{kang2021mlenhanced}}}
It has been observed that the performance of existing pure learned multi-dimensional indexes might degrade for high-dimensional data due to the ``curse of dimensionality". As a result, ML models have been incorporated into different traditional high-dimensional indexes (e.g., VA+Index~\cite{ferhatosmanoglu2000vaindex}, DS-tree~\cite{wang2013dstree}, iSAX~\cite{shieh2009isax}) to build the ML-enhanced variants of these traditional index structures. Notice that an R-tree has been used for implementing the VA+Index structure. Moreover, the proposed ML-enhanced indexes focus on improving recall for approximate kNN query processing in high-dimensional data. The underlying intuition behind all the ML-enhanced indexes is ``once an index is built, the distribution of the nearest neighbors given a query object has been fixed, and therefore is learnable". As a result, the proposed ML-enhanced indexes use deep neural networks to guide k-nearest-neighbor (KNN) search on traditional tree-based indexes. Particularly, the problem of kNN query processing has been formulated as a multi-class classification problem~\cite{aggarwal2015data_book}, where the goal is to predict the leaf nodes that contain the nearest neighbors given an input query. Hence, by leveraging the predictions of the ML models, the ML-enhanced indexes improve the leaf node access order of their traditional counterparts.

\paragraph{{\textbf{AI+R-tree~\cite{mamun2022ai+}}}}
Similar to the ML-enhanced index~\cite{kang2021mlenhanced}, the AI+R-tree incorporates ML models to enhance the performance of a traditional R-tree. As areas covered by R-tree nodes overlap in space, searching for a single object may require exploring multiple paths from root to leaf. This overlapping issue negatively impacts R-tree query processing performance. In the AI+R-tree, for a range query, an overlap ratio is proposed to quantify the degree of unnecessary leaf node accesses. Moreover, range query processing in an R-tree has been formulated as a multi-label classification problem~\cite{herrera2016multilabel}. Specifically, a new AI-tree is designed that trains multiple ML models to directly predict the true leaf node IDs for an input range query. To improve the query processing time of a traditional R-tree in the case of high-overlap queries, a hybrid AI+R-tree is proposed that processes high- and low-overlap queries using the AI-tree and the regular R-tree, respectively. The AI+R-tree uses the overlap ratio to train an ML model, enabling it to classify an input query as high- or low-overlap. It assumes that the data and query workloads are fixed. For efficient query processing, it also leverages a traditional grid structure to index the learned ML models~\cite{arefHandWritten1995}.

\paragraph{{\textbf{PolyFit}~\cite{li2021polyfit}}}
The PolyFit method has been proposed to process approximate range aggregate queries (e.g., SUM, COUNT). This method leverages piecewise polynomial functions for query processing, and constructs the index based on polynomial fitting for intervals of data. The choice of polynomial functions is due to their observed benefits over linear regression. Multiple polynomial functions are used to minimize the fitting error, as a single polynomial function might not fit the entire dataset accurately. Additionally, a greedy segmentation method has been proposed to reduce the number of polynomial functions. PolyFit is extended to support queries over multi-dimensional data by estimating cumulative functions for multiple keys and employing greedy segmentation techniques. However, due to the quadratic time complexity of the greedy approach in the multi-dimensional case, a Quad-tree~\cite{samet1984quadtree} has been incorporated to identify segments.

\paragraph{{\textbf{LearnedKD}~\cite{yongxin2020study}}}
LearnedKD integrates an ML model with a traditional k-d tree~\cite{Bentley1975kdtree} to speed up kNN query processing. For ML model training, kNN queries are initially processed using the traditional k-d tree, and the positions of the k nearest neighbors are recorded as labels. Then, a deep neural network model is trained on this dataset to assign high probabilities to data objects that are nearest neighbors of a given kNN query. This supervised learning process enables the model to predict whether an object is a nearest neighbor for an input query. However, the ML model's output does not directly identify the k nearest neighbors. Therefore, a post-processing step is necessary to determine the nearest neighbors after the prediction. This method is applied specifically to two-dimensional k-d trees.

\paragraph{{\textbf{LMI-1}~\cite{antol2021learned}}}
Inspired by the benefits of learned multi-dimensional indexes (e.g., Flood~\cite{nathan2020learning}), a Learned Metric Index (LMI-1, for short) has been proposed in~\cite{antol2021learned}. LMI-1 is specifically designed for indexing data in metric spaces, where similar data objects are clustered together using a distance metric. Moreover, LMI-1 suggests using predictions from a hierarchical structure with ML models for query processing rather than relying on their distance metric. Notice that LMI-1 is a hybrid structure that requires a pre-existing index structure to partition the data. Afterwards, it uses the partitions as labels to train a hierarchy of supervised ML models. LMI-1 is one of the earliest works to extend the concept of learned multi-dimensional indexes into the metric space.

\section{Learned Mutable Fixed Data Layout Indexes}~\label{section:Learning_Mutable_Fixed_one-d_multi-d}
In this section, we present the Mutable Fixed Data Layout Learned Indexes in both the one- and multi-dimensional spaces.
\subsection{The One-dimensional Case}

\subsubsection{\textbf{Pure Learned Indexes with In-Place Insertion Strategy}}
We present the core concepts of ALEX~\cite{ding2020alex}, a one-dimensional index that highlights the class of Mutable Pure Learned Indexes with Fixed Data Layout  and an In-place Insertion Strategy.

\paragraph{{\textbf{ALEX}~\cite{ding2020alex}}}
ALEX is an in-memory updatable learned index structure. Its core elements include: i)~RMI as the ML model hierarchical structure, and ii)~A Gapped Array (GA, for short) and A Packed Memory Array (PMA, for short)~\cite{bender2007adaptive} as leaf node layouts in RMI. ALEX adopts an adaptive RMI structure capable of handling the insertion of new keys. During initialization, the root model is executed first, partitioning the key space among its child nodes, that in turn are recursively initialized. Each non-root node is assigned a fixed number of partitions. When a partition size is appropriate, leaf nodes are created in the form of either GA or PMA to support inserts. The main idea behind the proposed GA is to maintain space/gaps at the leaf level so the structure can accommodate in-place ML model-based insertions. Moreover, if the partition size exceeds the maximum number of keys, a new inner node will be created. On the other hand, if the partition size is too small, adjacent partitions will be merged to avoid wasting leaf nodes. An extended version of ALEX, termed ALEX+, with concurrency support has been introduced in~\cite{wongkham2022updatable}.

Other indexes in this category are as follows. AIDEL~\cite{AIDEL} leverages independence among ML models to build a scalable index structure. TALI~\cite{guo2022tali} uses the update distribution of data for efficient lookup and insertion operations. StableBF~\cite{liu2020stable} designs a learned bloom filter for handling data streams. DriftModel~\cite{hadian2019considerations} addresses the issue of drift correction (i.e., model error due to updates) in updatable learned indexes. SALI~\cite{ge2023sali} employs probability models to construct a scalable index structure. APEX~\cite{lu2021apex} extends ALEX to develop an index optimized for persistent memory. CARMI~\cite{zhang2022carmi} offers a cache-friendly extension of the RMI index. EWALI~\cite{liu2022data} improves the write performance of a learned index by using dual buffers. TONE~\cite{zhang2022tone} reduces the tail latency of a learned index through a two-level leaf design. COLIN~\cite{zhang2021colin} is a cache-friendly learned index structure that leverages a mixture of learned and simple nodes (i.e., a heterogeneous node structure).

\subsubsection{\textbf{Pure Learned Indexes with Delta-Buffer Insertion Strategy}}
We present the core concepts of PGM~\cite{ferragina2020pgm}, a one-dimensional index that highlights the class of Mutable Pure Learned Indexes with Fixed Data Layout and a Delta-Buffer Insertion Strategy.

\paragraph{{\textbf{PGM}~\cite{ferragina2020pgm}}}
It has been observed that the initially proposed one-dimensional learned indexes (e.g., RMI) provide an empirical error bound without any formal worst-case bound. As a result, the Piecewise Geometric Model index (PGM, for short) is proposed with a guaranteed worst-case bound. The steps for constructing the PGM index are as follows: i)~Computing optimal piecewise linear segments with a pre-fixed model error bound $\epsilon$, ii)~Storing the segments as a triplet of key, slope, and intercept, iii)~Repeat the previous steps recursively. PGM 
uses
a delta buffer insertion strategy to support dynamic inserts. Moreover, three variants of PGM 
have been proposed: i)~Compressed PGM for space efficiency, ii)~Self-adaptive PGM for a particular query distribution, and iii)~Multi-criteria PGM 
that
optimizes 
for a user-given requirement, e.g., query time.

The other indexes in this category are as follows. ASLM~\cite{ASLM} employs simple single-layer ML models for efficient update handling. Doraemon~\cite{tang2019learned} re-uses pre-trained ML models for similar data distributions to reduce the model re-training cost. XIndex~\cite{tang2020xindex} supports concurrency natively by leveraging a two-phase compaction scheme. In ConcurrentLI~\cite{wang2022concurrent}, XIndex has been extended to XIndex-R and XIndex-H as range and hash indexes, respectively. Similar to XIndex, SIndex~\cite{wang2020sindex} is a concurrent learned index specifically designed for indexing strings. FINEdex~\cite{li2021finedex} avoids using a large delta buffer and employs instead a delta per training record to support data insertion. FILM~\cite{ma2022film} 
uses
a cold data identification mechanism 
that
enables efficient data swapping between disk and memory. DiffLex~\cite{cui2023difflex} is a NUMA-aware learned index that leverages sparse and dense arrays based on the hotness (i.e., cold vs. hot) of the keys. PLIN~\cite{zhang2022plin} utilizes Optimal Piecewise Linear Representation~\cite{xie2014maximum} to build an index optimized for non-volatile memory. 

\subsubsection{\textbf{Hybrid Learned Indexes}}
In this section, we present the core concepts of Bourbon~\cite{dai2020wisckey}, a One-dimensional that highlights the class of Hybrid Learned Indexes with Mutable Fixed Data Layout.

\paragraph{{\textbf{Bourbon}~\cite{dai2020wisckey}}}
Bourbon incorporates ML models to enhance the performance of an LSM-tree~\cite{luo2020lsm}. Moreover, Bourbon is an updatable ML-enhanced LSM-tree 
that
has been integrated inside WiscKey, a commercial key-value store~\cite{lu2017wisckey}. 
A
key observation 
in
Bourbon is that immutable SSTables are suitable for learning because there are no in-place updates. Particularly, the search index block component of the LSM-tree has been replaced with ML models. However, the learning of the SSTables has been categorized based on whether the SSTables are short- or long-lived. As a result, at runtime, a simple cost-benefit analyzer has been used to decide whether learning is beneficial. 

The other indexes in this category are as follows. TridentKV~\cite{lu2021tridentkv} adopts an optimized learned index block structure to build an ML-enhanced LSM-tree. SA-LSM~\cite{zhang2022sa} exploits a survival analysis technique in an LSM-tree based key-value store.
FITing-tree~\cite{GalakatosMBFK2019Fitingtree} combines a traditional B$^+$-tree with piecewise linear functions. AccB$^+$tree~\cite{llavesh2019accelerating} employs linear models to enhance the search operation of a traditional B$^+$-tree. MADEX~\cite{hadian2020madex} accelerates the intra-page lookup performance of a traditional B$^+$-tree by leveraging ML models. Sieve~\cite{tong2023sieve} incorporates piecewise linear models with a traditional B$^+$-tree to design an index for efficient block-skipping. LIFOSS~\cite{yu2023lifoss} exploits a two-layer learning model to design an index targeted for data streams. IFB-tree~\cite{hadian2019interp} designs interpolation-friendly nodes to enhance the performance of a traditional B-tree. Hybrid-LR~\cite{qu2019hybrid} exploits the benefit of linear regression models and B-trees to design a hybrid structure. 
PLBF~\cite{vaidya2020partitioned}, PLBF++~\cite{sato2023fast}, SNARF~\cite{vaidya2022snarf}, {IA-LBF}~\cite{bhattacharya2020adaptive}, PDDBF~\cite{zeng2023two}, BF-Sandwich~\cite{BF-sandwich10.5555/3326943.3326986}, and AdaBF~\cite{dai2019adaptive} combine ML models with  traditional bloom filters to build learned bloom filters.
FLIRT~\cite{yang2023flirt} exploits a circular queue structure to design a parameter-free index for data streams. 
HERMIT~\cite{wu2019designing} uses an ML-enhanced Tiered Regression Search Tree (TRS-Tree) to detect column correlations.
S3-SkipList~\cite{zhang2019s3} employs ML models to select guard entries in a traditional skip-list structure~\cite{pugh1990skip} .
RUSLI~\cite{mishra2021rusli} extends RS~\cite{kipf2020radixspline} to support data updates in real time.



\subsection{The Multi-dimensional Case}

\subsubsection{\textbf{A Summary of the Properties of the Mutable Fixed Data Layout Indexes}}
The properties of the class of mutable fixed data layout learned multi-dimensional indexes are presented in Table~\ref{table:fixed_layout_multi_d_summary_mutable}. We highlight the types of queries supported by each of the indexes. For point queries, some indexes do not explicitly provide a query processing algorithm or experimental results. However, if an index can easily be extended to support point queries, we have considered that the index supports point query processing. 
On the other hand, due to nature of the application domain, an index might be designed emphasizing efficiency over accuracy. As a result, an index can support either exact/approximate query processing. Here, approximate query processing refers to the event of missing some results from the set of exact answer. Here, we also indicate whether a supported query type returns an exact or an approximate answer.

\begin{table}[h!]
\caption{Properties of the Mutable Fixed Data Layout Learned Multi-dimensional Indexes} 
\centering 
\begin{tabular}{p{0.2\linewidth}p{0.07\linewidth}p{0.1\linewidth}p{0.1\linewidth}p{0.09\linewidth}p{0.07\linewidth}p{0.07\linewidth}p{0.05\linewidth}p{0.05\linewidth}} 
\toprule


Index&Data Layout&Pure Learned&Hybrid Learned&Data Space&Point Query&Range Query&kNN Query&Join Query\\
\midrule

GLIN-ALEX~\cite{wang2022glin} & Fixed & In-place  & $\times$ & Projected & Exact & Exact & $\times$ & $\times$\\
\hline
SLBRIN~\cite{wang2023slbrin} & Fixed & $\times$ & BRIN & Projected & Exact & Exact & Exact & $\times$\\
\hline
LSTI~\cite{ding2023learned} & Fixed & $\times$ & Radix Table & Projected & Exact & Exact & Exact & $\times$\\
\hline
IF-X~\cite{hadian2020handsoff} & Fixed & $\times$ & R-tree & Native & Exact & Exact & $\times$ & $\times$\\
\hline
Period Index~\cite{behrend2019period} & Fixed & $\times$ & Grid & Native & Exact & Exact & $\times$ & $\times$\\
\bottomrule
\end{tabular}
\label{table:fixed_layout_multi_d_summary_mutable}
\end{table}

\subsubsection{\textbf{Pure Learned Indexes with In-Place Insertion in the Projected Space}}
\paragraph{\textbf{GLIN-ALEX}~\cite{wang2022glin}}
Generic Learned Indexing (GLIN, for short) is a lightweight structure to index complex geometric objects (e.g., polygons). GLIN projects the multi-dimensional geometric objects into one-dimensional intervals of Z-values. After that, the objects are sorted based on their minimum Z-value. As a result, any existing order-preserving one-dimensional traditional or learned indexing technique can be applied to the sorted values. Here, GLIN uses both B-tree and ALEX~\cite{ding2020alex} indexes to 
produce
GLIN-BTREE and GLIN-ALEX, respectively. Moreover, for a particular leaf node, GLIN maintains the Minimum Bounding Rectangles (MBRs) of all objects of that node. As a result, during the ML model-based refinement step, a leaf node can be skipped if its MBR does not overlap with the query rectangle's MBR. GLIN can support \textit{containment} and \textit{intersect} queries for complex geometric objects. However, GLIN can have true negative results for \textit{intersect} queries. As a result, GLIN adopts a query augmentation technique to widen the Z-value interval. This query augmentation technique introduces a trade-off between correctness and pruning time. As ALEX adopts an in-place insertion strategy, GLIN-ALEX also uses the same strategy to support inserts.



\subsubsection{\textbf{Hybrid Learned Indexes in the Projected Space}}

\paragraph{\textbf{SLBRIN}~\cite{wang2023slbrin}}
Spatial Learned Index Based on BRIN (SLBRIN, for short) extends the ideas of Block Range Index (BRIN, for short)~\cite{yu2017indexing} in the context of learned spatial indexing. It aims to achieve high performance during query processing while maintaining efficient update operation. The main idea is to divide the index objects into two components: History Range (HR) and Current Range (CR). HR and CR are beneficial for query processing and update handling, respectively. For creating HR and index entries, SLBRIN first projects the multi-dimensional data into the one-dimensional space using Geohash~\cite{liu2014geohash}. This enables imposing a linear ordering using the Geohash values. After that, HR with index entries are partitioned recursively into ranges so that a one-dimensional learned index can be built for each range. SLBRIN employs Multi-layer Perceptron (MLP) as the ML model for training. Notice that CR
remains
empty during this index construction phase because there are no update operations at this step. During an update operation, the data will be encoded using Geohash to create the index entries. After that, the new index entries will be stored in CR for a short period of time. The new entries will be moved to HR during the periodic merge operation of CR. After a merge CR operation, a model re-training process is invoked to correct the error bound. SLBRIN can support point, range, and k-NN queries.

\paragraph{\textbf{LSTI}~\cite{ding2023learned}}
LSTI is a learned multi-dimensional index for processing spatio-textual data. Here, the multi-dimensional data is first projected into the one-dimensional space using Z/Morton-order~\cite{orenstein1984class, morton1966computer}. 
After that, techniques similar to the one-dimensional learned index RadixSpline~\cite{kipf2020radixspline} are applied to the projected data. The proposed index can support data updates using techniques similar to the one-dimensional updateable index ALEX~\cite{ding2020alex}. Moreover, it can process four types of queries: boolean point, boolean range, boolean kNN, and top-k text similarity. Given a specific text description. the boolean point, range, kNN, and top-k queries retrieve a specific location, the objects in a specific region, the objects closest to a specific location,  and multiple objects (with approximate text descriptions) in a specific spatial region, respectively.  
Notice that two parameters: max\_err and the number of radix bits significantly impact the index performance. As a result, a Random Forest~\cite{breiman2001random} regression model has been used to learn the optimal index parameters for specific data and query workloads.

\subsubsection{\textbf{Hybrid Learned Indexes in Native Space}}
Hybrid learned indexes combine traditional index structures with ML models to
build ML-enhanced index structures. Here, we present the hybrid learned indexes for the multi-dimensional space where these indexes operate in the native data space.

\paragraph{\textbf{Interpolation Friendly Indexes}~\cite{hadian2020handsoff}}
Interpolation Friendly Indexes (IF-X, for short) have been introduced in~\cite{hadian2020handsoff}. 
IF-X leverages the idea of augmenting traditional index structures (e.g., the R-tree) with ML models. Particularly, a lightweight technique,
e.g., linear interpolation~\cite{graefe2006b}, has been used for one of the dimensions. The use of a simple linear interpolation method offers the following benefits: faster computation time and fewer parameters. Moreover, to minimize interpolation error, the proposed index sorts the data entries on the most suitable dimension
that
is selected based on the least model prediction error. Furthermore, for performing interpolation, the structure of the leaf nodes of the index has also been modified so that the nodes can accommodate the additional information needed for the ML models. Notice that the proposed technique does not consider the query workload when selecting the sort dimension.

\paragraph{\textbf{Period Index}~\cite{behrend2019period} }
Period Index has been proposed to index intervals (i.e., temporal period data) by position and duration. A grid-based data structure with constant time lookup is utilized. The core idea of the proposed index can be divided into two parts: First, splitting the timeline into buckets of fixed size, where each bucket is further partitioned into cells (referring to the position of the interval); Second, arranging the cells in levels (referring to the duration of the interval). Moreover, it proposes finding the length of the buckets by adapting to the underlying distribution using a cumulative histogram. Hence, an improved version of the Period index, 
termed
Period Index*, has been proposed with this adaptive bucket length. 
Period Index* can be viewed as a learning-enhanced version of the Period Index. The proposed index can process range, duration, and range-duration queries.

\subsubsection{\textbf{Open Branches in the Taxonomy}}
As of 
the time of writing this article, 
there are no mutable learned multi-dimensional indexes in any of the following categories:
i)~Mutable Fixed Data Layout Multi-dimensional Pure Learned Indexes with In-place Insertion Strategy in Native Space, and ii)~Mutable Fixed Data Layout Multi-dimensional Pure Learned Indexes with Delta-Buffer Insertion Strategy. We envision that the above mentioned open branches will be useful for classifying learned multi-dimensional indexes in the future.



\section{Learned Mutable Indexes with Dynamic Data Layout  }~\label{section:Learning_Mutable_Dynamic_one-d_multi-d}
For the class of mutable learned indexes, if the layout of the data is arranged/re-arranged by the ML models while building the learned index, we refer to them as having a dynamic data layout. In this section, we present the class of Mutable Learned Indexes with Dynamic Data Layout in both the one- and the multi-dimensional spaces. 

\subsection{The One-dimensional Case}

\subsubsection{\textbf{Pure Learned Indexes with In-place Insertion Strategy}}
In this section, we present the core concepts of LIPP~\cite{LIPP10.14778/3457390.3457393} to highlight the class of Mutable   Pure Learned Indexes with with Dynamic Data Layout and In-place Insertion Strategy for the One-dimensional case.

\paragraph{{\textbf{LIPP}~\cite{LIPP10.14778/3457390.3457393}}}
It has been observed that the performance of an updatable one-dimensional learned index, 
e.g., 
ALEX~\cite{ding2020alex}, degrades in the presence of the ML model's misprediction. This degradation occurs because searching after an imprecise prediction always incurs overhead. To address this issue, the Learned Index with Precise Positions (LIPP, for short) has been proposed. The main idea behind LIPP is to avoid misprediction by ensuring that the key-to-position mapping is precise and eliminating the need for localized search following each misprediction. Consequently, the lookup cost is bounded by the tree height. Additionally, each LIPP node contains an ML model, a bit vector indicating entry type, and an array of entries. Each entry can be of the following types: NULL, DATA, or NODE. The NULL type indicates an unused slot, the DATA type represents a single entry, and the NODE type denotes a child node at the next level. Notably, LIPP does not distinguish between non-leaf and leaf nodes. During the construction phase of LIPP, if multiple keys map to the same position, a new child node is created. To ensure even distribution of mapping, kernelized linear models are utilized. Since LIPP is a sorted index, the selected kernel function must be monotonically increasing. The Fastest Minimum Conflict Degree (FMCD) algorithm is proposed for computing the model of each LIPP node. Furthermore, LIPP employs an in-place insertion strategy akin to ALEX. An enhanced version of LIPP supporting concurrency, LIPP+, has been introduced~\cite{wongkham2022updatable}.

The other indexes in this category are as follows. NFL~\cite{wu2022nfl} exploits a key distribution-transformation method so that the learned index can be built in the easy-to-learn transformed space. DILI~\cite{DILI10.14778/3598581.3598593} employs linear regression models to design a distribution-driven index structure. WIPE~\cite{wang2023wipe} is designed for reducing the issue of write-amplification in non-volatile memory.

\subsubsection{\textbf{Open Branches in the Taxonomy}}
As of now, there are no mutable learned one-dimensional indexes in any of the following categories in the one-dimensinoal case:
i) Mutable Pure Learned Indexes with Dynamic Data Layout and  Delta-Buffer Insertion Strategy, and ii) Mutable Hybrid Learned Indexes with Dynamic Data Layout.
We envision that the above mentioned open branches will be useful for classifying learned one-dimensional indexes in the future.


\subsection{The Multi-dimensional Case}


\subsubsection{\textbf{A Summary of the Properties of the Mutable Indexes with Dynamic Data Layout } }

\begin{table}[h!]
\caption{Properties of the Mutable Learned Indexes with Dynamic Data Layout in the Multi-dimensional Space} 
\centering 
\begin{tabular}{p{0.2\linewidth}p{0.07\linewidth}p{0.1\linewidth}p{0.1\linewidth}p{0.09\linewidth}p{0.07\linewidth}p{0.07\linewidth}p{0.06\linewidth}p{0.05\linewidth}} 
\toprule


Index&Data Layout&Pure Learned&Hybrid Learned&Data Space&Point Query&Range Query&kNN Query&Join Query\\
\midrule

LISA~\cite{LISA} & Dynamic & In-place & $\times$ & Projected & Exact & Exact & Exact & $\times$\\
\hline
RSMI~\cite{qi2020effectively} & Dynamic & In-place & $\times$ & Projected & Exact & Approx. & Approx. & $\times$\\
\hline
BMT~\cite{li2023towards} & Dynamic & In-place & $\times$ & Projected & Exact & Exact & Exact & $\times$\\
\hline
LMSFC~\cite{10.14778/3603581.3603598} & Dynamic & In-place & $\times$ & Projected & Exact & Exact & $\times$ & $\times$\\
\hline
LIMS~\cite{tian2022learned} & Dynamic & In-place & $\times$ & Native & Exact & Exact & Exact & $\times$\\
\hline
MTO~\cite{ding2021mto} & Dynamic & In-place & $\times$ & Native & Exact & Exact & $\times$ & Exact\\
\hline
WISK~\cite{sheng2023wisk} & Dynamic & Delta Buffer & $\times$ & Native & Exact & Exact & Exact & $\times$\\
\hline
LPBF~\cite{zou2022learned} & Dynamic & $\times$ & CBF & Projected & \NA & \NA & \NA & \NA\\
\hline
PA-LBF~\cite{zeng2023pa} & Dynamic & $\times$ & CBF & Projected& \NA & \NA & \NA & \NA\\
\hline
RLR-tree~\cite{gu2023rlr} & Dynamic & $\times$ & R-tree & Native & Exact & Exact & Exact & Exact\\
\hline
RW-tree~\cite{dong2022rw} & Dynamic & $\times$ & R-tree & Native & Exact & Exact & Exact & $\times$\\
\hline
ACR-tree~\cite{huang2023acr} & Dynamic& $\times$ & R-tree & Native & Exact & Exact & Exact & $\times$\\
\hline
PLATON~\cite{yang2023platon} & Dynamic & $\times$ & R-tree & Native & Exact & Exact & Exact & $\times$\\
\hline
Waffle~\cite{choi2022waffle} & Dynamic & $\times$ & Grid & Native & Exact & Exact & Exact & $\times$\\
\hline
FHSIE~\cite{su2023fast} & Dynamic& $\times$ & Grid & Native & Exact & Exact & Exact & $\times$\\

\bottomrule
\end{tabular}
\label{table:multi-d_summary_mutable_dynamic}
\end{table}

The properties of the class of mutable learned multi-dimensional indexes with dynamic data layout are presented in Table~\ref{table:multi-d_summary_mutable_dynamic}. 
In the table, 
we present the types of queries supported by each of the learned multi-dimensional indexes. For point queries, some indexes do not explicitly provide a query processing algorithm or experimental results. However, if an index can easily be extended to support point queries, we have considered that the index supports point query processing. Moreover, we also indicate whether a supported query type returns an exact or approximate answer. Note that we have excluded learned multi-dimensional bloom filters (e.g., LPBF~\cite{zou2022learned}, PA-LBF~\cite{zeng2023pa}) in the context of supported query types because they are considered probabilistic existence index structures.

\subsubsection{\textbf{Pure Indexes with In-place Insertion in the Projected Space}}
\paragraph{{\textbf{LISA}~\cite{LISA}}}
Considering the issues of storage consumption and high I/O cost of 
the 
R-Tree, 
LISA, a disk-based Learned Index for Spatial Data, 
has been introduced.
The main idea of LISA
is to generate a searchable data layout in disk pages for an arbitrary spatial dataset using a machine learning model. 
LISA
consists of four parts: i)~Space partitioning into a series of grid cells; ii)~A mapping function M for mapping search keys to 1D values; iii)~A monotonic prediction function SP
that
takes the output of M as input and predicts the shard ID; iv)~Local models for each shard for allocating, splitting, and merging pages. During query processing, given a query rectangle $qr$, it is decomposed into smaller rectangles. Then, the shard prediction function $SP$ is used to select the shard in each cell (overlapping with $qr$). Subsequently, local models are utilized to find the address of pages overlapping with $qr$. LISA supports range queries, KNN queries, insertions, and deletions. For KNN query processing, each KNN query is converted into a series of range queries. LISA supports dynamic insertions by employing a model-based in-place insertion policy.

\paragraph{\textbf{RSMI}~\cite{qi2020effectively}}
Recursive Spatial Model Index (RSMI, for short) is a disk-based updateable learned spatial index. Due to uneven gaps in the empirical CDF of the Z-order space-filling curve, RSMI leverages rank space-based ordering~\cite{qi2018theoretically} to project multi-dimensional points into the one-dimensional space. Data points are sorted in ascending order based on their one-dimensional projection values. Then, a set of points are packed into a block based on a predefined block size. To scale with larger datasets, RSMI recursively partitions the data and trains models for each partition. Moreover, multiple ML models are trained to learn the mapping (i.e., from search key to disk block ID) in the projected space. RSMI can support point, range, and kNN queries on point data. The proposed RSMI produces approximate answers with high accuracy. For exact query processing, RSMI advocates for a search operation similar to an R-tree. Note that all the proposed techniques are primarily focused on point data. 
Query processing performance may be negatively impacted if the proposed method is applied to data objects with extension. RSMI supports dynamic inserts by employing an in-place insertion policy.

\paragraph{\textbf{BMT}~\cite{li2023towards}}
It has been observed that each existing SFC has a pre-fixed projection function that neither considers the underlying data distribution nor the query workloads. As a result, a specific SFC cannot perform well across a variety of datasets and query workloads because a fixed Bit Merging Pattern (BMP, for short) is applied to the entire dataset to build a particular SFC (e.g., bit interleaving for Z-order). 
Thus, for a given data and query workloads, the core idea is to apply different BMPs to different subspaces to construct a Piecewise SFC. Hence, a tree-based structure has been designed where each leaf node refers to a subspace. Moreover, for each subspace (i.e., leaf node), the path from root to leaf creates the BMP by representing each internal node as a bit value for different dimensions. The proposed Bit Merging Tree (BMT, for short) 
preserves two important properties of an SFC: Injection  (i.e., generating a unique value for a given input), and Monotonicity~\cite{lee2007approaching}. The BMT construction process has been formulated as a sequential decision-making process so that reinforcement-learning-based (RL-based) techniques can be leveraged to learn an effective BMT construction policy. Particularly, a greedy policy has been incorporated into a Monte Carlo Tree Search (MCTS, for short)~\cite{browne2012survey}. The main idea of the greedy policy is to choose an action (i.e., choose a bit for each node from a pool of candidate bits) so that the RL agent can maximize its reward (e.g., query performance). Notice that executing queries on BMT to calculate the reward is a costly process. As a result, a new metric called ScanRange has been proposed to calculate the reward efficiently. The proposed BMT has been integrated into two previously proposed learned indexes as a replacement of a traditional SFC: BMT+RSMI~\cite{qi2020effectively} and BMT+ZM-index~\cite{wang2019learned}.

\paragraph{\textbf{LMSFC}~\cite{10.14778/3603581.3603598}}
Learned Monotonic Space Filling Curve (LMSFC, for short) has been proposed to learn a parameterized monotonic Z-order for given data and query workloads. The outcome of learning a parameterized Z-order is to find parameters for a particular instance to minimize query processing cost. To minimize the cost of query processing for a given instance, the problem of learning an optimal parameterized SFC is formulated as an optimization problem. A Bayesian optimization method termed SMBO~\cite{hutter2011sequential} is used to approximately solve the optimization problem. Then, an offline dynamic programming-based (and a heuristic-based approach) is proposed to pack data points into disk pages based on a density-based cost function. The goal 
is to reduce the dead space of the MBRs of the disk pages by packing the data points as tightly as possible. After packing the points into the pages, the minimum z-value from each page is used to create a sorted array. Then, a one-dimensional learned index (e.g., PGM~\cite{ferragina2020pgm}) is used to find the page (given a z-value as input). However, even with an instance-optimized Z-order SFC, the issue of false positives remains. As a result, recursive splitting of the query rectangle is proposed to further reduce the number of false positives during the search operation. The proposed LMSFC supports inserts by adopting an in-place insertion strategy.


\subsubsection{\textbf{Pure Indexes with In-place Insertion in Native Space}}


\paragraph{\textbf{LIMS}~\cite{tian2022learned}}
It has been observed that most existing techniques for learned multi-dimensional indexes (in vector space) may not be directly applicable for indexing in metric space due to the lack of coordinate structure and dimension information. Moreover, only the property of triangular inequality can be leveraged for pruning in a metric space. As a result, a Learned Index for Metric Space (LIMS, for short) has been proposed to efficiently support point, range, and kNN queries in the metric space. To achieve 
these
goals, LIMS clusters the data objects into multiple groups, and builds a learned index for each cluster. Additionally, it selects a set of data objects as pivots for each cluster so that the distance between the pivot and the data objects can be pre-computed. Notice that the pre-computation of distances in each cluster enables LIMS to build learned indexes on these distance values. It has been observed that the selection of pivots can significantly impact the performance during query processing.

\paragraph{\textbf{MTO}~\cite{ding2021mto}}
Multi-Table Optimizer (MTO, for short) is designed to reduce I/O cost for analytical workloads in multi-table datasets. Similar to the previously proposed Qd-tree~\cite{yang2020qdtree}, MTO 
leverages reinforcement learning to route each record to the corresponding data block, ensuring that all elements in one block are from the same table. Besides using simple predicates (e.g., A.X $<$ 100), MTO proposes using join-induced predicates to optimize the layout for all tables. For join-induced predicates, it evaluates subqueries in these predicates to obtain indexes that are candidates for join operations. Then, these predicates and the table are 
sent to the Qd-tree as input to train the model. During query processing, MTO uses the Qd-tree to guide block access. To support dynamic workloads, MTO adopts an in-place insertion policy while taking into account block access reduction and sub-tree reorganization cost in the reward function.


\subsubsection{\textbf{Pure Indexes with Delta Buffer Insertion in Native Space}}

\paragraph{\textbf{WISK}~\cite{sheng2023wisk}}
Workload-aware Learned Index for Spatial Keyword (WISK, for short) is a learned spatial index for keyword queries. Given data and query workloads, WISK employs ML models that can learn using both spatial and textual information. This is achieved in two phases: firstly, it learns an optimal data layout for a particular query workload, and builds the index based on the learned layout. The index construction steps are as follows: i)~ML models are trained to approximate the CDF of the spatio-textual objects, ii)~A cost estimation function is defined based on the learned CDF, iii)~SGD is applied to learn the optimal partition by minimizing the cost, iv)~The hierarchy of WISK is constructed using a bottom-up packing approach, v)~The packing process has been transformed into a sequential decision-making process so that it can be solved using a reinforcement learning technique. As the query processing cost of WISK largely depends on how the data objects are partitioned during the construction of the index, an optimal partitioning problem has been formulated to construct the leaf nodes of the WISK. It has been shown that the problem of partitioning with optimal cost is NP-hard. As a result, a heuristic algorithm using Stochastic Gradient Descent (SGD)~\cite{bottou1998online} has been proposed. WISK supports dynamic inserts by employing a delta-buffer insertion strategy.


\subsubsection{\textbf{Hybrid Learned Indexes in the Projected Space}}
\paragraph{\textbf{LPBF}~\cite{zou2022learned}}
Learned Prefix Bloom Filter (LPBF, for short) is a learned bloom filter that is particularly designed for spatial data. The core idea of the proposed LPBF is to project multi-dimensional spatial data into one-dimensional binary code using a space-filling curve (e.g., Z-order). After that, the prefixes of binary z-values are grouped into several clusters based on a pre-defined parameter. Moreover, the suffixes of the same prefix are learned using sub-Learned Bloom Filters (sub-LBF)~\cite{macke2018lifting}. LPBF 
leverages a traditional Counting Bloom Filter (CBF, for short)~\cite{fan2000summary} as a backup filter to support data updates. The use of CBF and sub-LBF helps 
minimize the false positive rate and the model training time for the proposed method.

\paragraph{\textbf{PA-LBF}~\cite{zeng2023pa}}
Prefix-Based and Adaptive Learned Bloom Filter (PA-LBF) is an extended version of the previously proposed LPBF~\cite{zou2022learned}. Notice that the concept of adaptation has been introduced 
during the process of sub-LBF model training. For each sub-LBF, the main extension is the use of an adaptive learning process to calculate the number of filter layers based on a pre-defined threshold. This adaptive learning approach enables PA-LBF to further minimize the false positive rate compared to the previously proposed LPBF.

\subsubsection{\textbf{Hybrid Learned Indexes in Native Space}}
\subsubsection{\textbf{RLR-tree}~\cite{gu2023rlr}}
During the traditional R-tree construction process, there are several existing node splitting strategies (e.g., linear, quadratic)~\cite{guttman1984r}. For a chosen node splitting strategy, the query processing performance will vary for different data and query workloads. Reinforcement Learning-based R-Tree (RLR-tree, for short) formulates the ChooseSubtree and Split operations of a traditional R-tree as a Markov Decision Process (MDP)~\cite{puterman1990markov} (i.e., sequential decision-making process). It 
employs RL techniques to learn models for the ChooseSubtree and Split operations. Notice that the RLR-tree does not need to modify the existing query processing algorithms; instead, it aims to construct a better R-tree by optimizing the ChooseSubtree and Split operation. As a result, it can support all the existing query processing algorithms designed for the R-tree. Moreover, the RLR-tree presents the design of state space, action space, transition, and reward for MDP formulation. For example, in the context of MDP formulation for ChooseSubtree, while inserting a new object, a state is a tree node with the following features: area increase, perimeter increase, overlap increase, and occupancy rate. For the action space, given a current state, the RL agent picks a child node into which the new entry will be inserted. For the transition, given a state and an action, the agent moves to a child node. If the child node is a leaf node, the agent reaches a terminal state. In the context of designing the reward function, the RLR-tree proposes using a reference tree with a pre-defined ChooseSubtree and Split strategy. As a result, the reward is calculated based on the difference between the query processing cost of the RLR-tree and the reference R-tree. Based on the above-mentioned MDP formulation, the RLR-tree adopts a Deep Q-Network learning method (DQN, for short)~\cite{mnih2015human} for training the agent.

\paragraph{\textbf{RW-tree}~\cite{dong2022rw}}
A traditional R-tree is constructed using one of the following methods: bulk loading vs. one-by-one insertion. In the context of one-by-one insertion, most existing techniques do not take the query workload into account while constructing the R-tree. As a result, a learning-based framework has been proposed for query workload-aware R-tree construction. 
The goal is to achieve better query processing performance for future queries generated from a similar query distribution. The proposed RW-tree is built on the following core concepts: learning the query workload distribution and leveraging a cost model for the accurate approximation of query execution time. For learning the workload distribution, in the pre-processing step, the entire space is partitioned into grid cells, and all the queries are mapped into their corresponding grid cell. After that, the statistics of the queries are collected per cell. Next, the grid cells are clustered based on queries with similar statistics. Given the learned workload distribution and a new data insertion choice, a cost model is proposed to approximately measure the query execution time. This enables the RW-tree to select the insertion strategy with the lowest cost. The RW-tree can process both range and k-NN queries.

\paragraph{\textbf{ACR-tree}~\cite{huang2023acr}}
It has been observed that existing methods for R-tree node packing can be categorized into the following types: i)~Heuristics to pack objects into parents in a bottom-up manner, or ii)~A greedy approach to partition nodes into child nodes in a top-down fashion. The goal of the existing methods is to optimize the R-tree construction for the short-term (i.e., without considering the long-term tree construction cost). Moreover, these heuristic-based or greedy methods also try to pack the nodes as full as possible. As the optimal R-tree node packing problem is NP-hard, an RL-based method has been proposed to construct the R-tree for optimizing the long-term tree cost. Particularly, the top-down R-tree construction methods have been formulated as an MDP~\cite{puterman1990markov}. 
Then,
an Actor-Critic~\cite{grondman2012survey} based RL method has been applied to construct the Actor-Critic R-tree (ACR-tree, for short). 
The Critic part is designated to estimate the long-term tree cost, and the Actor part is designated to make decisions (e.g., split or pack). The ACR-tree 
uses a grid-based method to better encode the spatial information. Moreover, based on the estimation of the long-term cost, it either splits a node or packs an object onto a single child node. Notice that the Actor-Critic training process is time-consuming. As a result, a bottom-up model training process with training sharing has been proposed to speed up the training process. The proposed ACR-tree supports both range and kNN queries.

\paragraph{\textbf{PLATON}~\cite{yang2023platon}}
Similar to the ACR-tree~\cite{huang2023acr}, it has been observed that both top-down and bottom-up R-tree packing methods cannot adapt to particular data and query workloads due to the use of fixed heuristics. Moreover, the top-down packing method adheres to a sub-optimal node partitioning policy by ignoring dependencies between partition decisions among different R-tree nodes. As a result, for given data and query workloads, a top-down Packing with Learned Partition (PLATON, for short) policy has been proposed to overcome the mentioned limitations. Notice that the problem of optimal R-tree packing has been proved to be NP-hard. As a result, the proposed node partitioning policy is learned by leveraging an RL-based MCTS~\cite{browne2012survey} technique. However, in the presence of a large state space and long action sequence, the existing MCTS algorithm suffers from slow convergence. Hence, a divide-and-conquer strategy has been proposed to reduce the size of the state space. Moreover, optimization techniques,
e.g.,
early termination and level-wise sampling, have been applied for faster convergence of the MCTS algorithm. During the R-tree construction phase, for particular data and query workloads, the top-down packing is performed based on the RL-based learned partitioning policy.

\paragraph{\textbf{Waffle}~\cite{choi2022waffle}}
Waffle is an in-memory grid-based indexing system for moving objects. Waffle is built on the following major components: grid index manager, lock manager, transaction manager, re-grid manager, and an online configuration tuning component. Notice that the proposed indexing system contains many different configuration knobs. As a result, the performance of Waffle heavily depends on the configuration of the knobs. Moreover, due to the dynamic nature of moving objects, the settings of different knobs require online adjustments. To 
address
this issue, an online configuration tuning component,  
termed
Wafflemaker, has been proposed. Wafflemaker is an RL-based component that suggests an optimized knob configuration for Waffle in an online fashion.

\paragraph{\textbf{FHSIE}~\cite{su2023fast}}
The Fast Hybrid Spatial Index with External Memory Support (FHSIE, for short) has been proposed to efficiently extend the concept of learned spatial indexing to secondary storage. FHSIE clusters spatial objects using unsupervised ML techniques based on a few parameters. During the index construction phase, the K-means clustering method is applied recursively to the spatial data to build a height-balanced hierarchical structure. 
Then, 
a grid structure is 
incorporated at the bottom level of the previously created hierarchical structure. The purpose of the grid structure is to accurately find the intersections of bottom-level clusters with an input query. For extending FHSIE to external memory, the main idea is to allocate disk blocks for each of the following components separately: i)~Internal models, ii)~Bottom-level models, and iii)~Cells of the grid. FHSIE 
supports inserts using an in-place insertion strategy. Although the proposed index can efficiently support point, range, and KNN queries in secondary storage, the index construction time can be significantly higher than that of traditional spatial indexes.

\subsection{Open Branches in the Taxonomy}
As of 
the time of writing this survey paper, 
there are no learned multi-dimensional indexes in the following category: Mutable Dynamic Data Layout Pure Learned Indexes with Delta-Buffer Insertion Strategy in Projected Space.
We envision that the above mentioned open branch will be useful for classifying learned multi-dimensional indexes in the future.  

\section{A Summary of ML Techniques for Learned Multi-dimensional Indexes}~\label{section:List of ML Techniques}
A summary of ML techniques used for learned multi-dimensional indexes is presented in Table~\ref{table:multi-d_ML_Models}.

\begin{table}[h!]
\caption{A Summary of ML Techniques for Learned Multi-dimensional Indexes} 
\centering 
\begin{tabular}{p{0.3\linewidth}p{0.6\linewidth}} 
\toprule
Index&ML Techniques\\
\midrule
LSTI~\cite{ding2023learned} & RadixSpline, Random Forest Regression\\
LISA~\cite{LISA} & Lattice Regression Model\\
LIMS~\cite{tian2022learned} & Ploynomial Regression\\
SageDB-LMI~\cite{kraska2019sagedb} & Linear Model\\
Tsunami~\cite{DingNAK2020tsunami} & Linear Model\\
COAX~\cite{hadian2023coax} & Linear Model\\
GLIN\_ALEX~\cite{wang2022glin} & Linear Model\\
Z-index~\cite{pai2022towards} & Density Model\\
FLOOD~\cite{nathan2020learning} & Piecewise Linear Model, RMI\\
PolyFit~\cite{li2021polyfit} & Polynomial Function\\
IF-X~\cite{hadian2020handsoff} & Linear Interpolation\\
SPRIG~\cite{Zhang2021sprig} & Spatial Interpolation Function\\
SPRIG+~\cite{zhang2022efficient} & Spatial Interpolation Function\\
Period-index*~\cite{behrend2019period} & Cumulative Histogram\\
LMI-2~\cite{slaninakova2021data} & K-means Clustering, Gaussian Mixture Model\\
FHSIE~\cite{su2023fast} & K-means Clustering\\
RW-tree~\cite{dong2022rw} & Density based Clustering\\
WAZI~\cite{pai2023workload} & Random Forest Density Estimation Model\\
LMSFC~\cite{10.14778/3603581.3603598} & Random Forest\\
``AI+R''-tree~\cite{mamun2022ai+} & Random Forest, Decision Tree\\
CaseSpatial~\cite{pandey2020case} & RadixSpline\\
Spatial-LS~\cite{pandey2023enhancing} & RadixSpline\\
Distance-bounded~\cite{ZacharatouKSPDM21} & RadixSpline\\
LPBF~\cite{zou2022learned} & Gradient Boosting\\
PA-LBF~\cite{zeng2023pa} & Gradient Boosting\\
ZM-index~\cite{wang2019learned} & Neural Network, Linear Model\\
LearnedKD~\cite{yongxin2020study} & Neural Network\\
LMI-1~\cite{antol2021learned} & Neural Network\\
ML-enhanced~\cite{kang2021mlenhanced} & Neural Network, Linear Model\\
CompressedBF~\cite{davitkova2021compressing} & Neural Network\\
LearnedBF~\cite{macke2018lifting} & Neural Network\\
ML-index~\cite{davitkova2020ml} & Neural Network\\
HM-index~\cite{wang2021spatial} & Neural Network\\
RSMI~\cite{qi2020effectively} & Neural Network\\
SLBRIN~\cite{wang2023slbrin} & Neural Network\\
QD-tree~\cite{yang2020qdtree} & Reinforcement Learning\\
MTO~\cite{ding2021mto} & Reinforcement Learning\\
RLR-tree~\cite{gu2023rlr} & Reinforcement Learning\\
Waffle~\cite{choi2022waffle} & Reinforcement Learning\\
BMT~\cite{li2023towards} & Reinforcement Learning\\
WISK~\cite{sheng2023wisk} & Reinforcement Learning\\
ACR-tree~\cite{huang2023acr} & Reinforcement Learning\\
PLATON~\cite{yang2023platon} & Reinforcement Learning\\
\bottomrule
\end{tabular}
\label{table:multi-d_ML_Models}
\end{table}

\section{Benchmarking Learned Indexes}~\label{section:Benchmarking}

In this section, 
we present the studies 
conducted 
for 
benchmarking of 
either the 
one- 
or 
the multi-dimensional learned indexes.
\subsection{The One-dimensional Case}
Several studies on benchmarking learned one-dimensional indexes are available in the literature.
In~\cite{bindschaedler2021towards}, 
several ideas have been proposed for designing new benchmarks that are better suited to evaluate learned systems. The Search On Sorted Data Benchmark (SOSD) has been presented in~\cite{Kipf2019SOSDAB, marcus2020benchmarking}, and 
However, 
SOSD only considers in-memory read-only workloads. Moreover, 
SOSD covers three learned indexes: RMI~\cite{kraska2018case}, RS~\cite{kipf2020radixspline}, and PGM~\cite{ferragina2020pgm}. A critical analysis of RMI has been presented in~\cite{maltry2022critical}.
This analysis investigates the impact of each hyperparameter of RMI on prediction accuracy, lookup time, and build time. Moreover, it provides guidelines to configure RMI.
In~\cite{andersen2022micro}, 
a micro-architectural analysis of ALEX has been presented. This study goes beyond high-level metrics (e.g., latency and index size) towards lower-level metrics (e.g., cache misses per cache level, average number of execution cycles). The experiments are based on Intel's Top-Down Micro-architecture Analysis Method (TMAM)~\cite{yasin2014top}.
Wongkham, et al.~\cite{wongkham2022updatable}, 
evaluate updateable learned indexes. Five one-dimensional learned indexes have been included in this study: ALEX, PGM, LIPP~\cite{LIPP10.14778/3457390.3457393}, XIndex~\cite{tang2020xindex}, and FINEdex~\cite{li2021finedex}. Moreover, this study proposes to extend ALEX and LIPP to support concurrency as ALEX+ and LIPP+.
In~\cite{sun2023learned},
a testbed has been developed to facilitate the design and testing of existing and upcoming learned index structures. It also presents the design choices of key components in learned indexes (e.g., insertion, concurrency, bulk loading). The experiments cover eight one-dimensional learned indexes.
In~\cite{lan2023updatable}, 
it has been shown how to extend four in-memory updatable one-dimensional learned indexes (e.g., Fitting-tree~\cite{GalakatosMBFK2019Fitingtree}, ALEX, PGM, LIPP) in a disk-based setting. An extensive experimental evaluation is presented for the proposed extensions.
In~\cite{ge2023cutting}, 
several one-dimensional updateable learned indexes (e.g., RMI, RS, ALEX, XIndex, FITing-tree, and PGM) have been evaluated along four criteria: approximation algorithm, index structure, insertion strategy, and model re-training strategy. Moreover, it also provides several guidelines for designing learned indexes.
The concept of using ML models as a replacement for a hash function 
has been 
introduced initially as a hash model index~\cite{kraska2018case}. In~\cite{sabek2021learned} and~\cite{sabek2022can},  
extensive experimental studies have been 
conducted
to analyze the performance of learned hash index structures. The main contribution of these studies is to provide insights on choosing a learned hash index versus a traditional hash index structure.

\subsection{The Multi-dimensional Case}
To the best of our knowledge, there is no comprehensive benchmarking on learned multi-dimensional indexes. As a result, it is still an important and open area of research. 

\section{Towards the Integration of Learned Indexes into Practical Systems}~\label{section:Towards Practical Systems}
A few studies have been 
conducting
on integrating learned index structures into practical systems. Although a few one-dimensional learned indexes have been successfully integrated into practical database/storage engines, similar efforts are still in progress in the context of learned multi-dimensional indexes. In this section, we discuss the studies that take a step forward towards the integration of learned indexes into practical systems.

\subsection{The One-dimensional Case}
In Google-index~\cite{abu2020learned}, a learned index has been integrated into a distributed, disk-based database system: Google's Bigtable~\cite{chang2008bigtable}. The integrated learned index has demonstrated significant improvements in read latency and memory footprint. In BOURBON~\cite{dai2020wisckey}, the proposed ML-enhanced LSM-tree has been incorporated into a 
production-quality system, and has achieved high performance compared with its traditional counterpart. In SA-LSM~\cite{zhang2022sa}, the proposed ML-enhanced LSM-tree has been incorporated into a commercial-strength LSM-tree storage engine, namely X-Engine. Sieve~\cite{tong2023sieve} has been integrated in a distributed SQL query engine, namely Presto~\cite{sethi2019presto}. Hybrid-LR~\cite{qu2019hybrid} has been implemented in PostgreSQL~\cite{stonebraker1990implementation}.

\subsection{The Multi-dimensional Case}
In ELSI~\cite{10184779}, a system for efficiently building and rebuilding learned 
multi-dimensional indexes has been proposed. ELSI supports any learned spatial index that projects the multi-dimensional points into one dimension (i.e., termed as projected space in this survey) to impose an ordering and process queries using the predict-and-scan method. Two index-building methods have been proposed based on space partitioning and reinforcement learning. Moreover, ELSI has been combined with four multi-dimensional indexes: ZM-index~\cite{wang2019learned}, ML-index~\cite{davitkova2020ml}, RSMI~\cite{qi2020effectively}, and LISA~\cite{LISA}. However, the integration of ELSI into a commercial database system (e.g., PostgreSQL) has been mentioned as future work. In SageDB~\cite{ding2022sagedb}, the ideas of the Qd-tree~\cite{yang2020qdtree} have been implemented for creating the data layout inside a practical system. In PLATON~\cite{yang2023platon}, the proposed ML-enhanced R-tree has been implemented on top of a real-world spatial library: libspatialindex~\cite{libspatialindex_online_ref}, 
and a practical database system: PostgreSQL $14.3$~\cite{postgresql_online_ref}, with the PostGIS\cite{PostGIS_online_ref} extension.



\section{Open Challenges and Future Research Directions}~\label{section:Open Challenges}
In this section, we highlight the open research challenges and future research directions in the context of learned multi-dimensional indexes.

\subsection{Total Ordering and Error Bound}
In the context of one-dimensional learned indexes, in most cases, the ML models are trained with the assumption that the underlying data is sorted with a total order. However, prediction-based ML models inherently do not provide any guarantee of accuracy when applied to unseen test data. As a result, if the ML models make an error during prediction, it needs to be corrected by a local search based on a predefined error bound. Notice that this error bound can be easily achieved in the case of one-dimensional data because we can sort the data (i.e., total ordering). However, this is not the case for multi-dimensional data because there is no total ordering in the multi-dimensional space. As a result, a class of learned multi-dimensional indexes projects the data into one-dimensional space to impose an ordering. Various space-filling curves have been used in this context, each with different advantages and limitations. On the other hand, the class of learned multi-dimensional indexes in 
native 
space does not employ a projection function. However, some learned multi-dimensional indexes might select one of the dimensions to impose an order in native space. Each of the learned multi-dimensional indexes in 
native
space, for exact query processing, has proposed a different mechanism for error correction. However, providing an error bound is challenging for the latter class of indexes, and further research addressing this aspect is needed.

\subsection{Choice of ML Models}
Due to recent advances in the area of machine learning, there are numerous types of ML models available with various levels of model complexity. In general, a complex model is expected to learn the underlying data distribution better than a simpler model. However, in the context of learned indexes, it is normally suggested to adopt a simple ML model (e.g., linear regression, decision tree) whenever possible. The main reason behind this suggestion is performance (e.g., less training time, smaller model size, low model prediction latency). Notice that traditional database indexes (e.g., B$^+$ tree, R-tree) are highly optimized data structures. As a result, a learned index should avoid using complex ML models so that the model building time and prediction latency do not become bottlenecks to achieving low index construction time and high query processing performance, respectively. However, it has also been observed that a customized ML model yields better performance than basic ML models. For example, in~\cite{eppert2021tailored}, a tailored regression model has been proposed to achieve better performance. Moreover, there is a growing interest in designing and using deep neural network-based ML models~\cite{amato2022suitability, amato2023neural, ferragina2023nonlinear} in the context of learned indexes. As a result, the choice of ML models needs to be further investigated for designing learned multi-dimensional indexes.

\subsection{Model Training and Re-training}
Training ML models is one of the most important parts of constructing learned indexes. Most of the proposed learned indexes try to minimize the model training time as much as possible. Moreover, changes in the underlying input data/query distribution should be detected as soon as possible, and a model re-training process should be triggered when necessary. Although these challenges are addressed in many of the proposed indexes, there is still ample room for future research in this area.

\subsection{Supporting Dynamic Inserts/Updates}
The initial learned indexes only considered read-only workloads in a static scenario. After that, new methods have been proposed to support insert/update operations.
However, supporting inserts/updates comes with the cost of periodic re-organization (e.g., re-training) of the learned index structures or incorporating mechanisms that require additional space (e.g., a gapped array). In this survey, we have mainly observed two main strategies to support dynamic inserts: In-place and Delta buffer. Although these challenges are addressed in the class of mutable multi-dimensional learned indexes, further research is needed to investigate the advantages and disadvantages of each insertion strategy.

\subsection{Concurrency}
In the context of designing learned indexes, for most cases, supporting concurrency has come as an afterthought. Only a few proposed methods (marked with an * symbol in the taxonomy~\ref{fig:taxonomy}) discuss the issue of concurrency and propose techniques to support concurrency natively. As a result, future research in this area should treat the issue of concurrency as a first-class citizen while designing or implementing learned indexes.

\subsection{Index Compression}
Both one- and multi-dimensional indexes have demonstrated significant benefits in terms of reduced storage requirements. Moreover, in the context of inverted indexes~\cite{zobel2006inverted}, the benefits of index compression by incorporating a learned one-dimensional index into an inverted index structure have been studied in~\cite{oosterhuis2018potential}. Additionally, the advantages of a learned multi-dimensional bloom filter for index compression are studied in CompressLBF~\cite{davitkova2021compressing}. 
However, exploring the potential benefits of index compression using learned multi-dimensional indexes in the context of other index types is an interesting direction for future research.

\subsection{Security}
The issue of security in the context of learned indexes is also little explored. The impact of poisoning attacks has been discussed in a recent study~\cite{kornaropoulos2022price}. It has been shown that ``every injection into the training dataset has a cascading impact on multiple data values." This attack can significantly reduce the performance of learned indexes. As a result, robust learned indexing methods should be designed to tackle issues related to security. Exploring this direction in the context of multi-dimensional learned indexes is an open research topic.

\subsection{Benchmarking}
Although several studies have been performed to benchmark the performance of one-dimensional learned indexes, a comprehensive benchmark on learned multi-dimensional indexes is still missing. There are many open-sourced multi-dimensional datasets available in various repositories (e.g., UCR-STAR~\cite{ghosh2019ucr}). However, collecting real query workloads for multi-dimensional and spatial data is challenging. As a result, many studies use real datasets but synthetic query workloads. A benchmarking study on multi-dimensional learned indexes with real data and query workloads will likely fill this gap.

\subsection{Theoretical Analysis}
There are a few theoretical studies~\cite{ferragina2020learned_theory, ferragina2021performance, zeighami2023distribution, chen2022learned} that mathematically analyze the reasons behind the performance gain of learned one-dimensional indexes over traditional indexes. In~\cite{ferragina2020learned_theory, ferragina2021performance}, it has been shown that learned one-dimensional indexes are provably better than traditional indexes. Moreover, it has been 
proven
that under certain assumptions, learned one-dimensional indexes can answer queries in $O(\log \log n)$ expected query time~\cite{zeighami2023distribution}. Furthermore, for the class of learned one-dimensional indexes adopting piece-wise linear segments, the theoretical analysis on the choice of the parameter $\epsilon$ has been presented in~\cite{chen2022learned}. As a result, theoretical analysis of 
the various 
components of learned multi-dimensional indexes is needed to better understand the benefits and limitations of existing techniques.

\subsection{Leveraging GPUs for Learned Indexes}
The benefit of natively implementing a learned one-dimensional index on a GPU has been presented in~\cite{zhong2022learned}. Particularly, the PGM index has been implemented on a GPU so that the index can exploit the high concurrency of the GPU. During query processing,  GPU-PGM achieves an order of magnitude performance gain over the CPU-PGM. Similar investigation in the context of learned multi-dimensional indexes is an interesting direction for future research.

\section{Conclusion}~\label{section:Conclusion}
In this survey, we provide an up-to-date coverage of the state-of-the-art in learned multi-dimensional indexes. We 
have introduced
a taxonomy to categorize both one- and multi-dimensional indexes using the following criteria: i)~Learning the index vs. Indexing the learned models, ii)~Immutable vs. mutable learned indexes, iii)~Mutable fixed data layout vs. Mutable dynamic data layout learned indexes, iv)~One-dimensional vs. multi-dimensional learned indexes, v)~Pure learned indexes vs. hybrid learned indexes, vi)~In-place vs. delta-buffer insertion strategy for learned indexes, and vii)~Projected vs. native space learned indexes. Moreover, we expect that 
future 
learned index structures can be placed in the 
introduced taxonomy 
in this paper 
based on the proposed criteria. We 
summarize the core ideas of $43$ learned multi-dimensional indexes, and highlight the similarities/differences of more than $60$ learned one-dimensional indexes. Furthermore, we have listed the supported query types and the underlying ML techniques for each of the learned multi-dimensional indexes. We 
include a timeline diagram 
that shows
the evolution of both one- and multi-dimensional learned index structures. Finally, we have discussed several open challenges and future research opportunities in the area of learned multi-dimensional indexes.

\section{Acknowledgements}
Walid G. Aref acknowledges the support of the National Science Foundation under Grant Number IIS-1910216.



\bibliographystyle{ACM-Reference-Format}

\appendix









\end{document}